\documentstyle[12pt]{article}

\begin{document}

\font\mybb=msbm10 at 12pt
\def\bb#1{\hbox{\mybb#1}}
\def\Z {\bb{Z}}
\def\R {\bb{R}}
 \def\unit{\hbox to 3.3pt{\hskip1.3pt \vrule height 7pt width .4pt \hskip.7pt
\vrule height 7.85pt width .4pt \kern-2.4pt
\hrulefill \kern-3pt
\raise 4pt\hbox{\char'40}}}
\def\II{{\unit}}
\def\cM {{\cal{M}}}
\def\half{{\textstyle {1 \over 2}}}

\begin{titlepage}

\begin{flushright}
UG-1/95 \\
QMW-PH-95-1\\
{\bf hep-th/9506156}\\
\end{flushright}

\begin{center}

\baselineskip20pt
{\large {\bf SOLUTION--GENERATING TRANSFORMATIONS}}
{\large {\bf AND THE}}
{\large {\bf STRING EFFECTIVE ACTION}}
\vspace{1cm}

\baselineskip15pt

{\bf Eric Bergshoeff\footnote{E-mail: {\tt bergshoe@th.rug.nl}}
and  Bert Janssen\footnote{E-mail: {\tt janssen@th.rug.nl}}}

{\it Institute for Theoretical Physics, University of Groningen}\\
{\it Nijenborgh 4, 9747 AG Groningen, The Netherlands}
\vspace{.5cm}

{\bf Tom\'as Ort\'{\i}n}\footnote{E-mail: {\tt
t.ortin@qmw.ac.uk} Address after October 1st 1995: CERN
Theory Division, CH-1211, Gen\`eve 23, Switzerland.}

{\it Department of Physics, Queen Mary \& Westfield College}\\
{\it Mile End Road, London E1 4NS, U.K.}
\end{center}


\begin{abstract}

We study exhaustively the solution-generating transformations
(dualities) that occur in the context of the low-energy effective action
of superstring theory.

We first consider target-space duality (``$T$~duality'')
transformations in absence of vector fields.  We find that for one
isometry the full duality group is $(SO^{\uparrow}(1,1))^{3}\times
D_{4}$, the discrete part ($D_{4}$) being non-Abelian.  We, then,
include non-Abelian Yang--Mills fields and find the corresponding
generalization of the $T$~duality transformations.  We study the
$\alpha^{\prime}$ corrections to these transformations and show that the
$T$~duality rules considerably simplify if the gauge group is embedded
in the holonomy group.

Next, in the case in which there are Abelian vector fields, we consider
the duality group that includes the transformation introduced by Sen
that rotates among themselves components of the metric, axion and vector
field.

Finally we list the duality symmetries of the Type~II theories with one
isometry.

\end{abstract}

\end{titlepage}


\newpage

\baselineskip15pt
\pagestyle{plain}


\section*{Introduction}

In recent years an active field of research has been the study of
modified Einstein equations.  The modifications that have been
considered consist in the addition to pure gravity of extra scalars,
antisymmetric tensor fields (called dilatons and axions, respectively)
and (Abelian or non-Abelian) Yang--Mills fields.  These modified
Einstein equations admit special solutions whose consistency crucially
depends on the presence of the new fields.  For examples of such new
solutions, see, for instance, the review articles
Refs.~\cite{Ca1,Ho1,Se2,Du1,Kh1} and references therein.

One motivation for studying the above-mentioned modifications to General
Relativity is that they arise in string theory.  In string theory
elementary particles are described as the excitations of a string rather
than as point-like objects.  The size of a string can be characterized
by a dimensionful pa\-ra\-me\-ter $\alpha^{\prime}$ (that can also be
understood as the inverse of the string tension) in such a way that, in
the so-called zero-slope limit $\alpha^{\prime} \rightarrow 0$, an
ordinary field theory of point particles is obtained.  This zero-slope
limit of string theory corresponds to a modified (or extended) Einstein
theory of the type discussed above.  The complete effective action
includes, at higher orders in $\alpha^{\prime}$, contributions which are
of higher order in the Riemann tensor and the Yang--Mills field
strength.  Since string theory claims to give a consistent description
of quantum gra\-vi\-ty, solutions of the string effective action are
expected to contribute to our understanding of quantum gravity.

Particularly interesting are solitonic and supersymmetric solutions
\cite{Ca1,Du1,Kh1} to the low-energy effective field theory since, for
different reasons, many of them are expected to be not just exact
solutions to the effective action to all orders in $\alpha^{\prime}$,
but exact solutions of string theory.

In general, it appears difficult to find exact solutions to the string
equations of motion.  One of the reasons for this is that knowledge
about the explicit form of the higher order $\alpha^{\prime}$
corrections to the string effective action have become available only
fairly recently \cite{Be1}.  Fortunately, if one con\-si\-ders spacetimes
with an isometry, there exist transformations which generate new
solutions out of old ones.  We will refer to all these symmetries of the
equations of motion as ``dualities''.

The ``target-space'' (``$T$~'') duality transformations of the Type~I
superstring effective action where first introduced in the bosonic
$\sigma$--model context for general backgrounds with one isometry by
Buscher in Refs.~\cite{Bu1} (see Ref.~\cite{Al1} for an updated review)
as discrete ($\Z_{2}$) transformations that interchange certain
components of the metric and axion fields.  Later Ro\v cek and Verlinde
\cite{Ro1} proved that when the orbit of the isometry is closed, the
backgrounds related by Buscher's transformation correspond to the
same  CFT.

This symmetry is also present in the zero-slope limit of the effective
action, and, in this context, (see Ref.~\cite{Gi1} for a review with
extensive references) the classical $T$~duality group was found to be
the continuous $O(1,1)_{\rm Sugra}$.

On the other hand, using string-field theory arguments, Sen found that
in presence of an additional Abelian vector field the duality symmetry
was bigger: $O(1,2;\Z)$ \cite{Se1}.  At the level of the classical
zero-slope effective action, there is a continuous $O(1,2;\R)_{\rm
Sugra}$ $T$~duality group.  The increase in symmetry is due to the fact
that we now can interchange certain components of the metric or axion
fields with certain components of the Abelian vector field.  We will
refer to this kind of transformations as ``Sen transformations''.

The necessity of isometries strongly suggests the use of techniques of
dimensional reduction and a close relationship between the duality
symmetries in the original dimensionality and the ``hidden symmetries''
of the dimensionally reduced theory \cite{Ma1}.  For supergravity
theories the hidden symmetry groups of most supergravity theories are
well known \cite{Cr1} and this has been a fruitful approach in the sense
that the duality groups of many dimensionally reduced theories have been
found (see, for instance, Ref.~\cite{Hu2} were this point of view is
advocated).  However, the relation with the symmetries of the
``original'' higher-dimensional theories has not always been carefully
studied.  It is our purpose to do this here, for the simple case of a
single isometry, distinguishing between those dualities which become
simple general coordinate transformations or gauge transformations in
higher dimensions and those which do not.  We will combine this study
with a thorough search for all discrete and continuous duality
transformations, relating them, when possible, to symmetries of the
$\sigma$-model or the Type~II theories.  We will consider three cases:
the (Type~I) superstring effective action (i) in absence of vector
fields, (ii) in presence of non-Abelian Yang--Mills fields and (iii) in
presence of a single Abelian vector field.

Our main results are:

\begin{enumerate}

\item We find more (both discrete and continuous) duality symmetries.
Since the situation in the literature is unclear, some of them were
(perhaps) known in different contexts and sometimes mistaken for each
other.  We clarify the situation.  In particular, we find
that the $T$~duality group of the Type~I theory with no vector fields
is $(SO^{\uparrow}(1,1))^{3}\times D_{4}$.  The appearance of this
finite non-Abelian group ($D_{4}$) is remarkable.

\item We generalize Buscher's (discrete) transformation to the case in
which there are non-Abelian Yang--Mills fields present. Any solution of
the zero-slope heterotic string theory effective action with one
isometry can now be ``$T$~dualized''.

\item We present the $\alpha^{\prime}$ corrections to the generalized
(discrete) Buscher's duality transformation and show that it becomes
considerably simpler if the gauge group is embedded in the holonomy
group.

\item We give the explicit form in terms of the higher-dimensional
theory fields of the finite Sen transformation (one Abelian vector field
present).

\item We list all the duality symmetries of the Type~II theories
(including eleven-dimensional supergravity) and relate them with each
other and with global coordinate transformations of the higher-dimensional
theories, when possible.

\item In this respect, we remark the fact that Buscher's discrete
duality transformation is an ``unexpected'' symmetry in the sense that
it is not a global coordinate transformation in higher
dimensions\footnote{If Buscher's discrete duality transformation
corresponded to a global coordinate transformation in higher
dimensions, it would be a symmetry of all theories which are obtained
from a higher dimensional one through dimensional reduction, which is
not true. Only theories with the ``right'' field content have this
symmetry.}. Then, from the higher-dimensional point of view, it is the
only interesting solution-generating transformation since all the other
transformations are then gauge transformations.

\end{enumerate}

We would like to stress that we are not going to perform full-fledged
compactifications, in the sense that in an expansion of the fields in
harmonic functions of the compact dimension we will only keep the
massless modes, {\it i.e.}~those with no dependence on the coordinate
that parametrizes the compact dimension.  The theories that we will
obtain in this way will effectively be lower-dimensional theories.  We
will refer to this procedure as {\it dimensional reduction}, to
distinguish it from (Kaluza-Klein) {\it compactification}.  Dimensional
reduction, which has traditionally and successfully been used in
supersymmetry and supergravity as a method to obtain new theories and
which has been used in many recent works on duality starting from
Ref.~\cite{Ma1}, will be enough for all of purposes.

As a matter of fact, we are ultimately interested in duality symmetries
of string theories.  While effective actions contain some information
about the string massless modes, at least enough to determine their
low-energy dynamics, they do not contain much information about the
massive modes.  It would not make any sense to study the Kaluza-Klein
massive modes (whose origin are the higher-dimensional massless modes)
without including the original string massive modes at the same time.
Dimensional reduction of effective actions is, then, not only the
simplest approach, but, in general, the only consistent approach from
the low-energy point of view\footnote{The only exception to this
conclusion might be eleven-dimensional {\it supergravity}, whose
Kaluza-Klein compactification on a circle seems to give the whole
spectrum of Type~IIA {\it superstring} theory \cite{Wi1}.}.

On the other hand, we expect that all duality symmetries of superstring
theories will  be duality symmetries of the effective field theories
(supergravity theories) \cite{Se3}. Then, the study of the duality
symmetries of effective actions is the easiest way to discover those of
the full string theory. In some cases, like the Type~II theories, where
it is not known how to include the Ramond-Ramond background fields in
the $\sigma$-model, it is also the only available way \cite{Be3}.

This being said, one should be aware that the effective theory does not
always give an adequate representation of the corresponding string
theory, particularly where non-perturbative in $\alpha^{\prime}$ effects
occur \cite{Al2}, the results obtained cannot be fully trusted and
should be understood as indications but never as proofs of the
corresponding results in string theory.  This is particularly important
in the case of Buscher's $T$~duality transformation.  It was shown in
Refs.~\cite{Ba1,Be2} that this transformation seems not to respect
unbroken spacetime supersymmetries.  This surprising effect has been
studied by different authors \cite{Ba2,Ha1,Al2} and the conclusion
seems to be that Buscher's duality transformation does {\it not} break
spacetime supersymmetry and that the usual representation of spacetime
supersymmetry (and hence the usual effective action) does not describe
correctly the dynamics of string theory in this limit.

The intrinsically {\it stringy} nature of this transformation as
different from the rest of the $T$~duality group shows itself here.
Since, as we will see, the rest of the conventional $O(1,1)_{\rm Sugra}$
$T$~duality group\footnote{As we have already said, and we will show in
the first section, the $T$~duality group is indeed bigger.} corresponds
to global coordinate reparametrizations in higher dimensions, it
respects automatically supersymmetry.

This article is organized as follows.  In Section~\ref{sec-DD-1} we
review the $T$~duality symmetries of the Type~I theory in the absence
of vector fields.  Here we dimensionally reduce the action in the
isometry direction, and we look for symmetries of the lower dimensional
theory, as advocated in Refs.~\cite{Ma1,Hu2}.  The purpose of this
section is to set up our notation and conventions, and to thoroughly
review the known results finding some new ones.

In Section~\ref{sec-vector} we use the technique of dimensional
reduction to find the generalization of the discrete (Buscher)
$T$~duality rules in the presence of non-Abelian vector fields.  We
discuss the relation between our results and the $\sigma$-model
description of $T$~duality.

In Section~\ref{sec-o12} we discuss how for the special case of an
Abelian vector field Sen's solution-generating transformation emerges.
In particular, we discuss the connection between the Sen transformation
at the one-hand-side and special general coordinate plus gauge
transformations at the other-hand-side.

Next, in Section~\ref{sec-alpha} we study the $\alpha^{\prime}$
corrections to the discrete (Buscher's) $T$~duality transformations.
Appendices~\ref{sec-ST} and~\ref{sec-SA} contain respectively the finite
form of Sen's duality transformations in $D$ dimensions and a review of
analogous results in the Type~II theories and in eleven-dimensional
supergravity. We also describe in this last Appendix the relation
between the duality transformations studied in the main body of the
paper and global coordinate transformations in higher dimensions.


\section{$T$~duality Without Vector Fields}
\label{sec-DD-1}

In this section we review the $T$~duality symmetries of the bosonic
sector of the zero-slope heterotic string effective action (which has
the same form as the bosonic string one).  Therefore, there are no vector
fields present.  To keep the discussion general we will work in $D$
dimensions specifying later, where necessary, to the case $D=10$.
Furthermore, for simplicity, we only assume the existence of one
isometry direction.  The results presented have an obvious
generalization to the case of several commuting isometries.

The $D$--dimensional action we start from is, in the zero-slope limit,
given by

\begin{equation}
S_{\rm{Sugra}}^{(D)}={\textstyle\frac{1}{2}}\int d^{D}x \sqrt{-\hat{g}}\
e^{-2\hat{\phi}} \left[ -\hat{R} +4(\partial\hat{\phi})^{2}
-{\textstyle \frac{3}{4}}\hat{H}^{2} \right]\, ,
\label{eq:actionD1}
\end{equation}

\noindent where the fields are the metric, the axion and the dilaton:

\begin{equation}
\left\{\hat{g}_{\hat{\mu}\hat{\nu}},\hat{B}_{\hat{\mu}\hat{\nu}},
\hat{\phi}\right\}\, ,
\end{equation}

\noindent and our conventions are those of Ref.~\cite{Be2}.  In
particular, the axion field-strength $\hat{H}$ is given by

\begin{equation}
\hat{H}_{\hat\mu\hat\nu\hat\rho} =
\partial_{[\hat\mu}\hat{B}_{\hat\nu\hat\rho]}\, .
\label{eq:H1}
\end{equation}

We are going to assume that all the backgrounds (solutions of this
theory) considered admit one isometry whose orbits can be parametrized
by the coordinate $x$, {\it i.e.}~we assume that there exists a Killing
vector $\hat{k}^{\hat{\mu}}$ such that the Lie derivative of all fields
with respect to $\hat{k}^{\hat{\mu}}$ is zero and such that

\begin{equation}
\hat{k}^{\hat{\mu}}\partial_{\hat{\mu}}=\partial_{\underline{x}}\, .
\end{equation}

\noindent It is natural to use {\it adapted coordinates}\footnote{All
the $D$--dimensional entities carry a hat and the $(D-1)$-dimensional
ones don't, and $\mu=\underline{0},\ldots,\underline{D-2}$;
$x=x^{\underline{D-1}}$.  To distinguish between curved and flat
indices, we underline the curved ones ($\xi^{\underline{x}}$, for
instance).} $(x^{\mu},x)$ such that all fields are independent of the
redundant coordinate $x$.  Then, the space splits into a
$(D-1)$-dimensional space parametrized by the coordinates $x^{\mu}$ and
an ``internal'' space parametrized by the coordinate $x$.  In this
internal space, by assumption, ``nothing happens'', there is no
dynamics, since the fields are independent of $x$.  The theory is
effectively $(D-1)$-dimensional, and therefore, following the point of
view advocated in Ref.~\cite{Ma1}, we will reduce dimensionally the
action Eq.~(\ref{eq:actionD1}) to find the corresponding effective
action.

First of all, in this coordinate system, the components of the Killing
vector are

\begin{equation}
\hat{k}^{\hat{\mu}} = \delta^{\hat{\mu} \underline{x}}\, ,
\hspace{.5cm}
\hat{k}_{\hat{\mu}}=\hat{g}_{\hat{\mu} \underline{x}}\, ,
\hspace{.5cm}
\hat{k}^{\hat{\mu}}\hat{k}_{\hat{\mu}}=
\hat{g}_{\underline{x}\underline{x}}\, .
\end{equation}

The zero-slope limit Buscher's $T$~duality rules were originally derived
using the two-dimensional $\sigma$-model approach in Refs.~\cite{Bu1}.
The explicit form of these transformations is:

\begin{equation}
\begin{array}{rclrcl}
\tilde{\hat{g}}_{\mu\nu} & = &
\hat{g}_{\mu\nu}-(\hat{g}_{\underline{x}\mu}\hat{g}_{\underline{x}\nu}-
\hat{B}_{\underline{x}\mu}\hat{B}_{\underline{x}\nu})
/\hat{g}_{\underline{x}\underline{x}}\, ,
\hspace{-1cm}
& & &
\\
& & & & &
\\
\tilde{\hat{B}}_{\mu\nu} & = &
\hat{B}_{\mu\nu}+(\hat{g}_{\underline{x}\mu}
\hat{B}_{\nu \underline{x}}-
\hat{g}_{\underline{x}\nu}\hat{B}_{\mu \underline{x}})
/\hat{g}_{\underline{x}\underline{x}}\, ,
\hspace{-1cm}
& & &
\\
& & & & &
\\
\tilde{\hat{g}}_{\underline{x}\mu} & = &
\hat{B}_{\underline{x}\mu}/\hat{g}_{\underline{x}\underline{x}}\, , &
\tilde{\hat{B}}_{\underline{x}\mu} & = &
\hat{g}_{\underline{x}\mu}/\hat{g}_{\underline{x}\underline{x}},
\\
& & & & &
\\
\tilde{\hat{g}}_{\underline{x}\underline{x}} & = &
1/\hat{g}_{\underline{x}\underline{x}}\, , &
\tilde{\hat{\phi}} & = & \hat{\phi} -\frac{1}{2}\log
|\hat{g}_{\underline{x}\underline{x}}|\, .
\end{array}
\label{eq:Buscher}
\end{equation}

The transformations Eqs.~(\ref{eq:Buscher}) also leave the zero-slope
limit action $S_{\rm Sugra}^{(D)}$, given in Eq.~(\ref{eq:actionD1}),
invariant in the sense that

\begin{equation}
S_{\rm{Sugra}}^{(D)}(\tilde{\hat{g}}, \tilde{\hat{B}}, \tilde{\hat{\phi}})=
S_{\rm{Sugra}}^{(D)}(\hat{g}, \hat{B},\hat{\phi}) +
\int d^{D} x\ A(\hat{g},\hat{B},\hat{\phi}) \partial_{\underline{x}}
B(\hat{g},\hat{B},\hat{\phi})\, ,
\label{eq:aaa}
\end{equation}

\noindent where $A$ and $B$ are some expressions in terms of $\hat{g},
\hat{B}$ and $\hat{\phi}$.  This pro\-per\-ty shows that target-space
duality is indeed a symmetry of the equations of motion (to this order
in $\alpha^{\prime}$) and therefore a solution-generating
transformation: if a configuration, independent of $x$, is a solution of
the zero-slope limit equations of motion, then the dual configuration
is also a solution.

A few remarks are in order

\begin{itemize}

\item The duality transformations (\ref{eq:Buscher}) are only
      well-defined if $\hat{g}_{\underline{xx}} \ne 0$. This is
      guaranteed by the condition that the Killing vector $\hat{k}$ is
      non-null.  For simplicity we consider from now on only
      the space-like case.  It is not difficult to generalize the
      formulae to the general case. It is remarkable that it is
      precisely the restriction of the Killing vector to be non-null
      that allows us to perform dimensional reduction in that direction.
      Things are radically different in the null case, and we do not
      know which kind of ``null duality symmetries'', if any, exist.
      Recently, the dimensional reduction of the Einstein theory in a
      null direction has been studied in Ref.~\cite{Ju1}, but it is not
      clear yet whether their results can be applied to our problem since
      in our case the existence of a null Killing vector is not enough
      to prove that the integrability condition $R_{vv}=0$ ($v$ is the
      corresponding null coordinate), on which their results are based,
      always holds.

\item For the special case in which the configuration is given by the
      product of a $(D-1)$-dimensional Minkowski space times a circle we
      have that $\hat{g}_{{\underline {xx}}} \sim R$, where $R$ is the
      radius of the torus, and the duality transformation corresponds to
      the well-known transformation $R \rightarrow 1/R$ \cite{Ki1}.

\item The dual of the dual gives back the original configuration.
      Therefore, this duality transformation, that we will call from now
      on $B$, generates a $\Z_{2}^{(B)}$ symmetry group.

\item We are after symmetries of the equations of motion\footnote{From
      the point of view of string theory, the only meaning of the
      effective action (when it exists) is that the conditions for the
      vanishing of the beta functionals can be derived from its
      minimization. From the supergravity point of view, the action is
      meaningful and a good symmetry of the theory will always leave
      the action invariant.}, and, therefore, we will consider as good
      symmetries transformations that, instead of leaving invariant the
      action, as $B$, scale it.

\end{itemize}

To show that Equation~(\ref{eq:aaa}) holds, it is convenient to use a
supergravity interpretation of duality via dimensional reduction
\cite{Ma1}.  We use the standard techniques of Scherk and Schwarz
in Refs.~\cite{Sc1}.  Thus, the $D$--dimensional fields decompose as
follows:

\begin{equation}
\begin{array}{rclrcl}
\hat{g}_{\mu\nu} & = &
g_{\mu\nu} -k^{2}A_{\mu}A_{\nu}\, , \hspace{1cm} &
\hat{B}_{\mu\nu} & = & B_{\mu\nu}+A_{[\mu}B_{\nu]}\, ,
\\
& & & & &
\\
\hat{g}_{\underline{x}\mu} & = & -k^{2}A_{\mu}\, , &
\hat{B}_{\underline{x}\mu} & = & B_{\mu}\, ,
\\
& & & & &
\\
\hat{g}_{\underline{x}\underline{x}} & = & -k^{2}\, ,&
\hat{\phi} & = & \phi +\frac{1}{2}\log k\, ,
\end{array}
\label{eq:DD-1}
\end{equation}

\noindent where

\begin{equation}
\left\{ g_{\mu\nu},B_{\mu\nu},\phi,A_{\mu},B_{\mu},k\right\}\, ,
\end{equation}

\noindent are the $(D-1)$-dimensional fields and we have used the
notation

\begin{equation}
k=|\hat{k}_{\hat{\mu}}\hat{k}^{\hat{\mu}}|^{\frac{1}{2}}\, .
\end{equation}

\noindent Observe that $\hat{k}_{\hat{\mu}}\hat{k}^{\hat{\mu}}= \hat
g_{\underline {xx}} = -k^{2}$.  Similarly, the $(D-1)$-dimensional
fields are given in terms of the $D$--dimensional fields by

\begin{equation}
\begin{array}{rclrcl}
g_{\mu\nu} & = & \hat{g}_{\mu\nu}-\hat{g}_{\underline{x}\mu}
\hat{g}_{\underline{x}\nu}/\hat{g}_{\underline{x}\underline{x}}\, ,&
B_{\mu} & = & \hat{B}_{\underline{x}\mu}\, ,
\\
& & & & &
\\
B_{\mu\nu} & = & \hat{B}_{\mu\nu}+
\hat{g}_{\underline{x}[\mu}\hat{B}_{\nu]\underline{x}}
/\hat{g}_{\underline{x}\underline{x}}\, ,\hspace{1cm}&
\phi & = & \hat{\phi}-\frac{1}{4}
\log|\hat{g}_{\underline{x}\underline{x}}|\, ,
\\
& & & & &
\\
A_{\mu} & = & \hat{g}_{\underline{x}\mu}
/\hat{g}_{\underline{x}\underline{x}}\, ,&
k & = & |\hat{g}_{\underline{x}\underline{x}}|^{\frac{1}{2}}\, .
\end{array}
\label{eq:D-1D}
\end{equation}

Ignoring the integral over $x$, the $D$--dimensional action
Eq.~(\ref{eq:actionD1}) is identically equal to\footnote{Dropping the
integral over $x$ is consistent with our dimensional reduction
philosophy.  However, when global transformations of this coordinate are
involved, we will find that the lower-dimensional action scales while
the higher-dimensional action is invariant.  In both dimensions the
equations of motion are invariant.}

\begin{eqnarray}
S_{\rm {Sugra(red)}}^{(D-1)} & = & {\textstyle\frac{1}{2}}
\int d^{(D-1)}x\  \sqrt{-g}\ e^{-2\phi}
\left[ -R +4(\partial\phi)^{2} -{\textstyle\frac{3}{4}} H^{2} \right.
\nonumber \\
& &
\nonumber \\
& &
\left. -(\partial\log k)^{2} +{\textstyle \frac{1}{4}}
k^{2}F^{2}(A) +{\textstyle \frac{1}{4}} k^{-2}F^{2}(B)\right] \, ,
\label{eq:actionD-11}
\end{eqnarray}

\noindent where

\begin{eqnarray}
F_{\mu\nu}(A) & = & 2\partial_{[\mu}A_{\nu]}\, ,
\hspace{1.5cm}
F_{\mu\nu}(B)  =  2\partial_{[\mu}B_{\nu]}\, ,
\nonumber \\
& &
\nonumber \\
H_{\mu\nu\rho} & = & \partial_{[\mu}B_{\nu\rho]}+
{\textstyle \frac{1}{2}}A_{[\mu}F_{\nu\rho]}(B)
+{\textstyle \frac{1}{2}}B_{[\mu}F_{\nu\rho]}(A)\, .
\end{eqnarray}

The action Eq.~(\ref{eq:actionD-11}) can be interpreted as a
$(D-1)$-dimensional action for the above $(D-1)$-dimensional fields.  It
is known since the old supergravity days \cite{Cr1,Ga1} that this action
is invariant under the rigid non-compact (``supergravity'') symmetry
group

\begin{equation}
O(1,1)_{\rm Sugra} = SO^{\uparrow}(1,1)_{\rm Sugra} \ \times\
\Z_{2}^{(B)} \times\ \Z_{2}^{(S)}\, .
\end{equation}

\noindent $SO^{\uparrow}(1,1)$ is the part of $O(1,1)$ connected with
the identity, and $\Z_{2}^{(B)}\times\ \Z_{2}^{(S)}$ is its {\it mapping
class group}.  The first of the discrete symmetries, $\Z_{2}^{(B)}$, is
generated by the transformation that we denote by $B$:

\begin{equation}
\tilde{A}_{\mu} =   B_{\mu}\, ,\hspace{1.2cm}
\tilde{B}_{\mu}  =  A_{\mu}\, ,\hspace{1.2cm}
\tilde{k}^{2}  =  k^{-2}\, ,
\label{eq:dualreduced}
\end{equation}

\noindent while the other fields are invariant.  In $D$ dimensions these
rules correspond to Buscher's duality rules Eqs.~(\ref{eq:Buscher})
(hence the name).  The second $\Z_{2}$ symmetry is generated by the
transformation

\begin{equation}
A^{\prime}_{\mu} =  -A_{\mu}\, ,\hspace{1.2cm}
B^{\prime}_{\mu} =  -B_{\mu}\, ,
\label{eq:S}
\end{equation}

\noindent that we denote by $S$.  On the other hand, the
$SO^{\uparrow}(1,1)_{\rm Sugra}$ transformation (with continuous rigid
parameter $\alpha$)\footnote{The parameter $\alpha$ takes values in
$\R$.  Therefore, $SO^{\uparrow}(1,1)$ is isomorphic to the
multiplicative group $\R^{+}$ or to the additive group $\R$.} is a
scaling of the fields with diferent powers (``weights'') of the factor
$e^{\alpha}$.  The weights in nine dimensions ({\it i.e.}~when we take
$D=10$) are
given in Table~\ref{tab:weights-I}.

As it is explained in Appendix~\ref{sec-SA}, and following the classical
reasoning of Ref.~\cite{Cr1}, since the action above was obtained by
dimensional reduction of the coordinate $x$, one would generally expect
an $SO(1,1)$ $T$~duality group of re\-sca\-lings and reflections of the
compact coordinate $x^{\prime}=e^{\alpha}x,\ x^{\prime}=-x$.  This is
exactly the origin of $SO^{\uparrow}(1,1)_{Sugra}\times\Z_{2}^{(S)}$,
and an analogous global symmetry is expected in any theory which can be
obtained from (one)-dimensional reduction of another theory.  It will
come as no surprise that in all the cases considered in the following
sections this duality group is always present and we will not mention
its origin again.  On the other hand, the presence of the second
$\Z_{2}^{(B)}$ is completely unexpected (for instance, it is not present
in Type~II actions as explained in Ref.~\cite{Be3}) and ultimately
related to the stringy origin of the action we started with.

\begin{table}
\begin{center}
\begin{tabular}{||c||c|c|c|c|c|c|c||}
\hline\hline
& & & & & & & \\
Name & $g_{\mu\nu}$ & $B_{\mu\nu}$ & $A_{\mu}$ & $B_{\mu}$ & $e^{\phi}$
& $k$ & $S^{(9)}_{\rm Sugra(red)}$ \\
\hline\hline
& & & & & & & \\
$SO^{\uparrow}(1,1)_{\rm Sugra}$ &0&0&1&-1&0&-1&0\\
\hline
& & & & & & & \\
$SO^{\uparrow}(1,1)_{x-y}$ &1&1&1&0&$\frac{7}{4}$&$-\frac{1}{2}$&0\\
\hline
& & & & & & & \\
$SO^{\uparrow}(1,1)_{\rm string}$ &0&0&0&0&1&0&-2\\
\hline\hline
\end{tabular}
\end{center}
\caption{\textit{Weights of the heterotic fields under the two
$SO^{\uparrow}(1,1)$ symmetries of the action $S_{\rm Sugra (red)}^{(9)}$ and
the third $SO^{\uparrow}(1,1)$ which scales it in nine dimensions
(so taking $D=10$).}}
\label{tab:weights-I}
\end{table}

In addition to these well-known symmetries, there is another
$SO^{\uparrow}(1,1)_{x-y}$ scaling transformation and a discrete
$\Z^{(A)}_{2}$ transformation that leave the action invariant.  The
weights of the fields under the $SO^{\uparrow}(1,1)_{x-y}$ scaling in
nine dimensions are given in the second row in
Table~\ref{tab:weights-I}.

The $\Z^{(A)}_{2}$ symmetry group is generated by the transformation
that we call $A$

\begin{equation}
B_{\mu\nu}\rightarrow -B_{\mu\nu}\, ,\hspace{1.2cm}
B_{\mu}\rightarrow -B_{\mu}\, .
\label{eq:A}
\end{equation}

\noindent One may naively think that the total symmetry group is just
$SO^{\uparrow}(1,1)_{\rm Sugra}\times\ SO^{\uparrow}(1,1)_{x-y}\times\
\Z^{(B)}_{2} \times\ \Z^{(S)}_{2} \times\ \Z^{(A)}_{2}$.
However, a careful analysis shows that the actual symmetry group is

\begin{equation}
SO^{\uparrow}(1,1)_{\rm Sugra}\times\ SO^{\uparrow}(1,1)_{x-y}\times\
D_{4}\, .
\end{equation}

\noindent where $D_{4}$ is the symmetry group of rotations of a square
with undirected sides, which has two generators $b,c$ that obey
\footnote{See for instance Ref.~\cite{Jo1} where a representation in
terms of two-dimensional matrices is given in page $25$.}
$c^{4}=b^{2}=(bc)^{2}=1$.  In our case the generators are $b=B$
(Buscher's duality transformation Eq.~(\ref{eq:Buscher})) and the order
four element $c=AB$ (the $A$ is given by Eq.~(\ref{eq:A})).  Note that
$A$ and $B$ do not commute.

Finally we note that there is an additional $SO^{\uparrow}(1,1)_{\rm
string}$ scaling transformation (``string'') which is a symmetry of the
equations of motion but not of the action, which scales under it.  The
non-zero scaling weights in nine dimensions are given in the last line
of Table~\ref{tab:weights-I}.

We conclude that the full group of global symmetries of the equations of
motion is

\begin{equation}
SO^{\uparrow}(1,1)_{\rm Sugra}\times\
SO^{\uparrow}(1,1)_{x-y}\times\
SO^{\uparrow}(1,1)_{\rm string}\times\
D_{4}\, .
\end{equation}

The whole symmetry group (except for the $B$ transformation) can be
understood from a higher dimensional point of view and from the Type~II
theories point of view.  This is discussed in Appendix~\ref{sec-SA}.
Also, all the transformations in the discrete part of the symmetry group
$D_{4}$ can be understood from the $\sigma$-model with a $D$-dimensional
target-space point of view.  In particular, the transformation $A$
consists in the change of sign of the $D$-dimensional axion
($\hat{B}_{\hat{\mu}\hat{\nu}}\rightarrow
-\hat{B}_{\hat{\mu}\hat{\nu}}$) plus the interchange between
right-movers and left-movers $z\rightleftharpoons \overline{z}$.


\section{Duality In Presence Of Non-Abelian Vector Fields}
\label{sec-vector}

Since our ultimate goal is to study the duality symmetries of the full
heterotic string effective action, it is natural to study, as an
intermediate step, the effect of the addition of non-Abelian vector
fields on the duality symmetries found in the previous section.  Then,
our starting point is the so-called ``Sugra+YM'' action:

\begin{equation}
S^{(D)}_{\rm Sugra+YM} = {\textstyle\frac{1}{2}}\int d^{D}x
\sqrt{-\hat{g}}\ e^{-2\hat{\phi}} \left[ -\hat{R}+
4(\partial\hat{\phi})^{2}- {\textstyle \frac{3}{4}} \hat{H}^{2}
+{\textstyle\frac{1}{4g^{2}}} \hat{F}^{I}_{\hat{\mu}\hat{\nu}}
\hat{F}_{I}^{\hat{\mu}\hat{\nu}} \right]\, ,
\label{eq:actionvector}
\end{equation}

\noindent in terms of the fields

\begin{equation}
\left\{\hat{g}_{\hat{\mu}\hat{\nu}},\hat{B}_{\hat{\mu}\hat{\nu}},
\hat{V}^{I}_{\hat{\mu}},\hat{\phi}\right\}\, ,
\end{equation}

\noindent which, in the case $D=10$ corresponds to the bosonic sector of
$N=1,D=10$ supergravity coupled to Yang--Mills and is interesting by
itself.  Here $\hat{F}^{I}$ is the curvature of the Yang--Mills vector
field $\hat{V}^{I}$, $I$ is a Yang--Mills index (which we raise and
lower with $\delta_{IJ}$) and $g$ is the coupling constant. $\hat{H}$
contains now a Yang--Mills Chern--Simons term:

\begin{eqnarray}
\hat{F}_{\hat\mu\hat\nu}^{I} (\hat{V}) & = &
2\partial_{[\hat{\mu}} \hat{V}_{\hat{\nu}]}^{I}
-f_{KL}{}^{I} \hat{V}_{\hat{\mu}}^{K} \hat{V}_{\hat{\nu}}^{L}\, ,
\nonumber \\
& &
\nonumber \\
\hat{H}_{\hat{\mu}\hat{\nu}\hat{\rho}}  & = &
\partial_{[\hat{\mu}}\hat{B}_{\hat{\nu}\hat{\rho}]}
-{\textstyle\frac{1}{2g^{2}}} \left[
\hat{V}^{I}_{[\hat{\mu}}\hat{F}_{\hat{\nu}\hat{\rho}]I}(\hat{V})
+{\textstyle \frac{1}{3}} f_{IJK}\hat{V}^{I}_{[\hat\mu}
\hat{V}^{J}_{\hat{\nu}} \hat{V}^{K}_{\hat{\rho}]} \right]\, .
\label{eq:H}
\end{eqnarray}

In principle, there is an ambiguity in the relative sign between
$\partial \hat{B}$ and the Yang--Mills Chern--Simons term.  In fact,
there are two theories whose only difference is this relative sign and
which are related by the change of sign of
$\hat{B}_{\hat{\mu}\hat{\nu}}$ ($\Z_{2}^{(A)}$) which is no longer a
symmetry of each separate theory.  We, therefore, anticipate that the
group $D_{4}$ is broken to $\Z^{(B)}_{2}\times\ \Z_{2}^{(S)}$ in each
theory.  In fact $\Z_{2}^{(A)}$ is a duality transformation that brings
us from one theory to the other, exactly as happens in the Type~II
duality of Ref.~\cite{Be3}.  From the $\sigma$-model point of view,
these theories are related by a change of the sign of
$\hat{B}_{\hat{\mu}\hat{\nu}}$ and the simultaneous interchange of left-
and right-movers.  For the sake of definiteness, we will work with the
above choice of relative sign.

Following the standard procedure for dimensional reduction \cite{Sc1}
(see also Section~\ref{sec-DD-1}) we get in the $(D-1)$-dimensional
theory one graviton, one axion, two Abelian vector fields, one
non-Abelian vector field, the dilaton, a scalar and a set of additional
scalars that transform in the adjoint representation of the Yang--Mills
group:

\begin{equation}
\left\{g_{\mu\nu}, B_{\mu\nu}, A_{\mu}, B_{\mu}, V_{\mu}^{I}, \phi, k,
\ell^{I}\right\}\, .
\end{equation}

\noindent These fields are related to the $D$--dimensional fields by

\begin{eqnarray}
g_{\mu\nu} & = &
\hat{g}_{\mu\nu}-\hat{g}_{\underline{x}\mu} \hat{g}_{\underline{x}\nu}
/\hat{g}_{\underline{x}\underline{x}}\, ,
\nonumber \\
& &
\nonumber \\
B_{\mu\nu} & = &
\hat{B}_{\mu\nu}+\hat{g}_{\underline{x}[\mu}
\hat{B}_{\nu]\underline{x}} /\hat{g}_{\underline{x}\underline{x}}
-{\textstyle\frac{1}{2g^{2}}}\hat{V}^{I}_{\underline{x}}
\hat{g}_{\underline{x}[\mu}
\hat{V}_{\nu]I} /\hat{g}_{\underline{x}\underline{x}}\, ,
\nonumber \\
& &
\nonumber \\
V^{I}_{\mu} & = &
\hat{V}^{I}_{\mu}-\hat{V}^{I}_{\underline{x}}
\hat{g}_{\underline{x}\mu}/\hat{g}_{\underline{x}\underline{x}}\, ,
\hspace{1cm}
A_{\mu} =
\hat{g}_{\underline{x}\mu} /\hat{g}_{\underline{x}\underline{x}}\, ,
\nonumber \\
& &
\nonumber \\
B_{\mu} & = &
\hat{B}_{\underline{x}\mu} -
{\textstyle\frac{1}{2g^2}}
\hat{V}^I_{\underline{x}} \hat{V}_{\mu I}
+{\textstyle\frac{1}{2g^2}}
\hat{V}^{I}_{\underline{x}}\hat{V}_{\underline{x}I}
\hat{g}_{\underline{x}\mu}/\hat{g}_{\underline{x}\underline{x}}\, ,
\nonumber \\
& &
\nonumber \\
\phi & = &
\hat{\phi} -{\textstyle \frac{1}{4}} \log
|\hat{g}_{\underline{x}\underline{x}}|\, , \hspace{1cm}
k  =  |\hat{g}_{\underline{x}\underline{x}}|^{\frac{1}{2}}\, ,
\hspace{1cm}
\ell^{I} =  {\textstyle\frac{1}{g}}\hat{V}^{I}_{\underline{x}}\, .
\label{eq:fieldsvector}
\end{eqnarray}

\noindent The $D$--dimensional fields decompose into the
$(D-1)$-dimensional ones as follows:

\begin{equation}
\begin{array}{rclrcl}
\hat{g}_{\mu\nu} & = & g_{\mu\nu} -k^{2}A_{\mu}A_{\nu}\, ,
\hspace{.5cm}&
\hat{B}_{\mu\nu} & = & B_{\mu\nu}+A_{[\mu}B_{\nu]}
+\frac{1}{g}\ell^{I}A_{[\mu}V_{\nu]I}\, ,
\\
& & & & &
\\
\hat{g}_{\underline{x}\mu} & = & -k^{2}A_{\mu}\, , &
\hat{B}_{\underline{x}\mu} & = & B_{\mu}+\frac{1}{2g}\ell^{I}V_{\mu I}\,
, \\
& & & & &
\\
\hat{g}_{\underline{x}\underline{x}} & = & -k^{2}\, , &
\hat{\phi} & = & \phi +{\textstyle \frac{1}{2}}\log k\, ,
\\
& & & & &
\\
\hat{V}^{I}_{\underline{x}} & = & g\ \ell^{I}\, , &
\hat{V}^{I}_{\mu} & = & V^{I}_{\mu}+g\ \ell^{I}A_{\mu}\, .
\end{array}
\end{equation}

\noindent The $(D-1)$-dimensional vector fields and two-form are
defined in such a way that they transform in a standard way specified
below under the (infinitesimal) gauge symmetries that they inherit from
the $D$--dimensional fields:

\begin{enumerate}

\item $x$-independent reparametrizations $x^{\prime}=x-\xi(x^{\mu})$:

\begin{equation}
\begin{array}{rclrcl}
\delta \hat{g}_{\mu\nu} & = &
2\hat{g}_{\underline{x}(\mu} \partial_{\nu)}\xi\, ,
\hspace{1.2cm}&
\delta \hat{B}_{\mu\nu} & = &
-2\hat{B}_{\underline{x}[\mu}\partial_{\nu]} \xi\, ,
\\
& & & & &
\\
\delta \hat{g}_{\mu \underline {x}} & = &
\hat{g}_{\underline{xx}}\partial_{\mu}\xi\, , &
\delta \hat{V}^{I}_{\mu} & = & \hat{V}^{I}_{\underline{x}}
\partial_{\mu}\xi\, ,
\end{array}
\end{equation}

\item $x$--independent gauge transformations of the axion field:

\begin{equation}
\delta \hat{B}_{\hat{\mu}\hat{\nu}} =
2\partial_{[\hat{\mu}}\hat{\Sigma}_{\hat{\nu}]}\, ,
\end{equation}

\item $x$--independent gauge transformations of the Yang--Mills field
      (accompanied by a Nicolai--Townsend (N--T) transformation of the
      axion two-form)

\begin{eqnarray}
\delta \hat{V}^{I}_{\hat{\mu}} & = &
\partial_{\hat{\mu}}\Lambda^I + f_{JK}{}^{I}
\Lambda^{J}\hat{V}^{K}_{\hat{\mu}}\, ,
\nonumber \\
& &
\nonumber \\
\delta \hat{B}_{\hat{\mu}\hat{\nu}} & = &
{\textstyle\frac{1}{g^{2}}}\hat{V}^{I}_{[\hat{\mu}}
\partial_{\hat{\nu}]} \Lambda_{I}\, .
\end{eqnarray}
\end{enumerate}

These three gauge symmetries correspond to the following four gauge
symmetries of the $(D-1)$-dimensional fields:

\begin{enumerate}

\item Gauge transformations of the vector field $A_\mu$ (plus N--T
      transformation of the axion two-form):

\begin{eqnarray}
\delta A_{\mu} & = & \partial_{\mu}\xi\, ,
\nonumber \\
& &
\nonumber \\
\delta B_{\mu\nu} & = & - B_{[\mu} \partial_{\nu]}\xi\, ,
\end{eqnarray}

\item Gauge transformations of the axion field

\begin{equation}
\delta B_{\mu\nu} = 2 \partial_{[\mu} \Sigma_{\nu]}\,
\end{equation}

\noindent where $\Sigma_{\mu}=\hat{\Sigma}_{\mu}$,

\item Gauge transformations of the vector field $B_\mu$ (plus a N--T
      transformation)

\begin{eqnarray}
\delta B_{\mu} & = & \partial_{\mu}\Sigma\, ,
\nonumber \\
& &
\nonumber \\
\delta B_{\mu\nu} & = & -A_{[\mu} \partial_{\nu]} \Sigma\, ,
\end{eqnarray}

\noindent where $\Sigma=-\hat{\Sigma}_{\underline{x}}$,

\item Gauge transformation of the vector field $V_\mu$ (plus a N--T
      transformation)

\begin{eqnarray}
\delta V_{\mu}^{I} & = & \partial_{\mu}\Lambda^{I}
+ f_{JK}{}^{I}\Lambda^{J} V^{K}_{\mu}\, ,
\nonumber \\
& &
\nonumber \\
\delta B_{\mu\nu} & = & {\textstyle\frac{1}{g^{2}}}
V^{I}_{[\mu}\partial_{\nu]} \Lambda_{I}\, .
\end{eqnarray}

\end{enumerate}

The gauge--invariant $(D-1)$-dimensional vector and axion field
strengths are, accordingly

\begin{eqnarray}
F_{\mu\nu}(A) & = & 2\partial_{[\mu}A_{\nu]}\, ,
\hspace{1.5cm}\,
F_{\mu\nu}(B) =  2\partial_{[\mu}B_{\nu]}\, ,
\nonumber\\
& &
\nonumber\\
F^{I}_{\mu\nu}(V) & = &  2\partial_{[\mu}V^{I}_{\nu]} -
f_{JK}{}^{I} V_{\mu}^{J} V_{\nu}^{K}\, ,
\nonumber \\
& &
\nonumber\\
H_{\mu\nu\rho} & = & \partial_{[\mu}B_{\nu\rho]}+
{\textstyle \frac{1}{2}} A_{[\mu}F_{\nu\rho]}(B)
+{\textstyle \frac{1}{2}} B_{[\mu}F_{\nu\rho]}(A)
\nonumber\\
& &
\nonumber\\
& &
-{\textstyle\frac{1}{2g^{2}}}\left[ V^{I}_{[\mu}F_{\nu\rho]I}(V)
+{\textstyle\frac{1}{3}}f_{IJK}V_{[\mu}^{I} V_{\nu}^{J}
V_{\rho]}^{K}\right]\, .
\end{eqnarray}

The dimensionally reduced action is given by

\begin{eqnarray}
S^{(D-1)}_{\rm Sugra+YM (red)} & = &
{\textstyle\frac{1}{2}}\int d^{(D-1)}x\ \sqrt{-g}\ e^{-2\phi}
\left\{
\vphantom{\frac{2k^{2} +\ell^{2}}{4k^{2}}}
-R +4(\partial\phi)^{2} -{\textstyle \frac{3}{4}}H^{2}\right.
\nonumber\\
& &
\nonumber\\
& & +{\textstyle\frac{1}{4g^{2}}}\left( \frac{k^{2}
+\ell^{2}}{k^{2}}\right) {\rm Tr}\ F^{2}(V)
-\left[(\partial\log k)^{2}+\frac{1}{2k^{2}}({\cal D} \ell)^{2} \right]
\nonumber \\
& &
\nonumber \\
& & +{\textstyle\frac{1}{4}}\left[\frac{(2k^{2}
+\ell^{2})^{2}}{4k^{2}}F^{2}(A)
+k^{-2}F^{2}(B)+\frac{\ell^{2}}{k^{2}}F(A)F(B)\right]
\nonumber \\
& &
\nonumber \\
& & \left. +{\textstyle\frac{1}{g}}F^{I}(V)
\left[\ell_{I}\left( \frac{2k^{2} +\ell^{2}}{4k^{2}} \right) F(A)
+\frac{\ell_{I}}{2k^{2}}F(B)\right]\right\}\, ,
\label{eq:actionvectorreduced}
\end{eqnarray}

\noindent where $\ell^{2} \equiv \ell^{I}\ell_{I}\, ,\,\, ({\cal
D}\ell)^{2} \equiv {\cal D}_{\mu}\ell^{I}{\cal D}^{\mu}\ell_{I}$
and the covariant derivative ${\cal D}_{\mu} \ell^{I}$ is defined by

\begin{equation}
{\cal D}_{\mu} \ell^{I} = \partial_{\mu} \ell^{I} +
f_{JK}{}^{I}\ell^{J}V_{\mu}^{K}\, .
\end{equation}

As happened in the previous section, the action is invariant under a
rigid $SO^{\uparrow}(1,1)_{\rm Sugra}\times\ \Z^{(B)}_{2} \times\
\Z^{(S)}_{2}$ symmetry.  The continuous $SO^{\uparrow}(1,1)_{\rm Sugra}$
transformations are scalings and the weights of the fields in nine
dimensions are given in Table~\ref{tab:weights-IYM}.

\begin{table}
\begin{center}
\begin{tabular}{||c||c|c|c|c|c|c|c|c|c|c||}
\hline\hline
& & & & & & & & & &\\
Name & $g_{\mu\nu}$ & $B_{\mu\nu}$ & $A_{\mu}$ & $B_{\mu}$ & $e^{\phi}$
& $k$ & $\ell$ & $V_{\mu}$ & $1/g^{2}$ & $S^{(9)}$ \\
\hline\hline
& & & & & & & & & &\\
$SO^{\uparrow}(1,1)_{\rm Sugra}$ &0&0&1&-1&0&-1&-1&0&0&0\\
\hline
& & & & & & & & & & \\
$SO^{\uparrow}(1,1)_{x-y}$
&1&1&1&0&$\frac{7}{4}$&$-\frac{1}{2}$&$-\frac{1}{2}$&0&1&0\\
\hline
& & & & & & & & & & \\
$SO^{\uparrow}(1,1)_{\rm string}$ &0&0&0&0&1&0&0&0&0&-2\\
\hline\hline
\end{tabular}
\end{center}
\caption{\textit{Weights of the {\rm Sugra+YM} fields under the two
$SO^{\uparrow}(1,1)$ (pseudo-) duality symmetries of the action
$S_{\rm Sugra+YM(red)}^{(9)}$ and the third $SO^{\uparrow}(1,1)$ which
scales it in $D=9$.}} \label{tab:weights-IYM}
\end{table}

The action of the discrete $B$ transformation that generates the first
$\Z^{(B)}_{2}$ is

\begin{equation}
\begin{array}{rclrcl}
\tilde{A}_{\mu}
&
=
&
B_{\mu}\, , \hspace{2cm}
&
\tilde{B}_{\mu}
&
=
&
A_{\mu}\, ,
\\
& & & & &
\\
\tilde{k^{2}}
&
=
&
\frac{4k^{2}}{(\ell^{2}+2k^{2})^{2}}\, , \hspace{1.2cm}
&
\tilde{\ell}^{I}
&
=
&
\frac{2\ell^{I}}{\ell^{2}+2k^{2}}\,
\\
\end{array}
\label{eq:dualreducedvector1}
\end{equation}

\noindent and the action of the transformation $S$ which generates
$\Z_{2}^{(S)}$ is

\begin{equation}
\begin{array}{rclrcl}
A^{\prime}_{\mu}
&
=
&
-A_{\mu}\, , \hspace{2cm}
&
B^{\prime}_{\mu}
&
=
&
-B_{\mu}\, ,
\\
& & & & &
\\
\ell^{I\prime}
&
=
&
-\ell^{I}\, .
& & &
\\
\end{array}
\label{eq:dualreducedvector2}
\end{equation}

If one allows for transformations of the coupling constant $g$ ({\it
pseudo-dualities} \cite{Hu1}) the $SO^{\uparrow}(1,1)_{x-y}$ symmetry of
the action found in the previous section can be promoted to a symmetry
in the presence of Yang--Mills fields.  The weights are given in the
second row of Table~\ref{tab:weights-IYM} for nine dimensions.  Finally,
the $SO^{\uparrow}(1,1)_{\rm string}$ trivially extends to this case.
In consequence the full symmetry group of the equations of motion is

\begin{equation}
SO^{\uparrow}(1,1)_{\rm Sugra} \times\ SO^{\uparrow}(1,1)_{x-y} \times\
SO^{\uparrow}(1,1)_{\rm string} \times\ \Z_{2}^{(B)} \times\ \Z_{2}^{(S)}.
\end{equation}

\noindent We see that the non-Abelian discrete group $D_4$ from the
previous section indeed breaks into the Abelian group $\Z_2^{(B)}
\times\ \Z_2^{(S)}$, as mentioned above, due to the presence of the
Yang--Mills Chern--Simons term in the axion field strength.

\noindent In $D$ dimensions the transformation $B$ corresponds to the
following ge\-ne\-ra\-li\-za\-tion of Buscher's $T$~duality
rules\footnote{The analogous result for Abelian vector fields was first
given in Refs.~\cite{Gi2,Sha1}}:

\begin{equation}
\begin{array}{rclrcl}
\tilde{\hat{g}}_{\mu\nu} & = &
\hat{g}_{\mu\nu}+\left[ \hat{g}_{\underline{x}\underline{x}}
\hat{G}_{\underline{x}\mu}\hat{G}_{\underline{x}\nu}
-2\hat{G}_{\underline{x}\underline{x}}\hat{G}_{\underline{x}(\mu}
\hat{g}_{\nu)\underline{x}}\right]
/\hat{G}_{\underline{x}\underline{x}}^{2}\, ,\hspace{-4cm}
\\
& & \\
\tilde{\hat{B}}_{\mu\nu} & = &
\hat{B}_{\mu\nu}-\hat{G}_{\underline{x}[\mu}
\hat{G}_{\nu]\underline{x}}/\hat{G}_{\underline{x}\underline{x}}\, ,
\\
& & \\
\tilde{\hat{g}}_{\underline{x}\mu} & = & -\hat{g}_{\underline{x}\mu}
/\hat{G}_{\underline{x}\underline{x}}
+\hat{g}_{\underline{x}\underline{x}}\hat{G}_{\underline{x}\mu}
/\hat{G}_{\underline{x}\underline{x}}^{2}\, , &
\tilde{\hat{B}}_{\underline{x}\mu} & = &
-\hat{B}_{\underline{x}\mu}/\hat{G}_{\underline{x}\underline{x}}
+\hat{G}_{\underline{x}\mu}/\hat{G}_{\underline{x}\underline{x}}\, ,
\\
& &
\\
\tilde{\hat{g}}_{\underline{x}\underline{x}} & = &
\hat{g}_{\underline{x}\underline{x}}
/\hat{G}_{\underline{x}\underline{x}}^{2}\, , &
\tilde{\hat{\phi}} & = & \hat{\phi}-\frac{1}{2}
\log |\hat{G}_{\underline{x}\underline{x}}|\, ,
\\
& &
\\
\tilde{\hat{V}}^{I}_{\underline{x}} & = & -\hat{V}^{I}_{\underline{x}}
/\hat{G}_{\underline{x}\underline{x}}\, , &
\tilde{\hat{V}}^{I}_{\mu} & = &
\hat{V}^{I}_{\mu} -\hat{V}^{I}_{\underline{x}}\hat{G}_{\underline{x}\mu}
/\hat{G}_{\underline{x}\underline{x}}\, ,
\end{array}
\label{eq:dualvector}
\end{equation}

\noindent with

\begin{equation}
\hat{G}_{\hat{\mu}\hat{\nu}}=
\hat{g}_{\hat{\mu}\hat{\nu}}
+\hat{B}_{\hat{\mu}\hat{\nu}}
-{\textstyle\frac{1}{2g^{2}}}\hat{V}^I_{\hat{\mu}}
\hat{V}_{\hat{\nu}I}\, .
\label{eq:Ghat}
\end{equation}

These are the duality rules of the theory corresponding to the choice of
sign in Eq.~(\ref{eq:H}).  Since the theory corresponding to the other
choice can be obtained by performing an $A$ transformation
($\hat{B}_{\hat{\mu}\hat{\nu}} \rightarrow
-\hat{B}_{\hat{\mu}\hat{\nu}}$), its duality rules can be also obtained
by performing an $A$ transformation in the above rules.

Note that the factor $\sqrt{-\hat{g}}\ e^{-2\hat{\phi}}$ which occurs as
an overall factor in the $D$--dimensional Lagrangian is invariant under
this set of transformations, since the determinant of the metric
transforms as follows:

\begin{equation}
\sqrt{-\tilde{\hat{g}}}\ =\ \hat{G}_{\underline{x}\underline{x}}^{-1}\
\sqrt{-\hat{g}}\, .
\end{equation}

Finally, it is remarkable that the duality rules of
$\hat{G}_{\hat{\mu}\hat{\nu}}$ are of the following particular simple
form:

\begin{equation}
\begin{array}{rclrcl}
\tilde{\hat{G}}_{\mu\nu} & = &
\hat{G}_{\mu\nu}-\hat{G}_{\underline{x}\mu}\hat{G}_{\nu \underline{x}}
/\hat{G}_{\underline{x}\underline{x}}\, ,
\hspace{1.2cm}&
\tilde{\hat{G}}_{\underline{x}\underline{x}} & = &
1/\hat{G}_{\underline{x}\underline{x}}\, ,
\\
& &
\\
\tilde{\hat{G}}_{\underline{x}\mu} & = &
+\hat{G}_{\underline{x}\mu}/\hat{G}_{\underline{x}\underline{x}}\, , &
\tilde{\hat{G}}_{\mu \underline{x}} & = &
-\hat{G}_{\mu \underline{x}}/\hat{G}_{\underline{x}\underline{x}}\, .
\end{array}
\label{eq:simple}
\end{equation}


\section{Duality In Presence Of One Abelian Vector Field}
\label{sec-o12}

A particularly interesting case of the action considered in the previous
section is the one in which the gauge group is $\left( U(1)\right)^{n}$,
because new duality transformations that interchange components of the
metric or axion with components of the Abelian vector fields are now
possible\footnote{The Abelian case is obtained by first rescaling $V
\rightarrow gV$ and then putting all structure constants equal to zero.
Observe that after this rescaling, the weight of $V$ under
$SO^{\uparrow}(1,1)_{x-y}$ becomes $\frac{1}{2}$ and the
``pseudo-duality'' becomes a duality.}.  It is known that, as a
consequence, the $O(1,1)_{\rm Sugra}$ duality group of
Section~\ref{sec-vector} becomes $O(1,n+1)_{\rm Sugra}$
\cite{Sha1,Gi2,Ga1}.  The symmetries $SO^{\uparrow}(1,1)_{x-y}\times\
SO^{\uparrow}(1,1)_{\rm string}$ do not extend to larger symmetry groups
and remain as in the previous section.

We are only going to consider the case $n=1$ since it is the simplest
and shows all the interesting features.

The $(D-1)$-dimensional action is again given by
Eq.~(\ref{eq:actionvectorreduced}), specified to the Abelian case.  This
action can be rewritten into the following form, presented in
Refs.~\cite{Ma1,Ga1}, which makes the $O(1,2)_{\rm Sugra}$--invariance
manifest :

\begin{eqnarray}
S^{(D-1)}_{{\rm Sugra}+U(1)} & = &
{\textstyle\frac{1}{2}}\int d^{(D-1)}x\ \sqrt{-g}\ e^{-2\phi}
\left\{-R +4(\partial\phi)^{2} -{\textstyle \frac{3}{4}}H^{2} \right.
\nonumber\\
& &
\nonumber\\
& & \left. +{\textstyle \frac{1}{8}}{\rm Tr}\ \left( \partial_{\mu}
M^{-1}\partial^{\mu} M \right) - {\textstyle \frac{1}{4}}
{\cal F}^i_{\mu\nu}({\cal A}) {\cal F}^{\mu\nu}_{i}({\cal A})
\right\}\, ,
\end{eqnarray}

\noindent where

\begin{eqnarray}
{\cal A}^i_\mu & = & \pmatrix{ A_\mu\cr
                                B_\mu\cr
                                V_\mu}\, ,
\hspace{1.2cm}
{\cal F}^{i}_{\mu\nu}({\cal A}) = \partial_{\mu} {\cal A}_{\nu}^{i} -
\partial_{\nu} {\cal A}_{\mu}^{i}\, ,
\nonumber \\
&&
\\
{\cal F}_{\mu\nu i} & = & M^{-1}_{ij} {\cal F}_{\mu\nu}^{j}\, ,
\hspace{.4cm} i=1,2,3\,
\nonumber
\end{eqnarray}

\noindent and

\begin{equation}
M^{-1}_{ij} =
\left(
\begin{array}{ccc}
-(2k^{2}+\ell^{2})^{2}/4k^{2} & -\ell^{2}/2k^{2} &
-(2k^{2}\ell+\ell^{3})/2k^{2} \\
& & \\
-\ell^{2}/2k^{2} & -1/k^{2} & -\ell/k^{2} \\
& & \\
-(2k^{2}\ell+\ell^{3})/2k^{2} & -\ell/k^{2} & -(k^{2}+\ell^{2})/k^{2} \\
\end{array}
\right) \, .
\end{equation}

\noindent The explictly $O(1,2)_{\rm Sugra}$--invariant axion
field-strength can be written as

\begin{equation}
H_{\mu\nu\rho} = \partial_{[\mu}B_{\nu\rho]} +{\textstyle \frac{1}{2}}
{\cal A}^{i}_{[\mu} {\cal F}^{j}_{\nu\rho]}({\cal A})\eta_{ij}\, ,
\end{equation}

\noindent where $\eta$ is the $O(1,2)$ metric in an off-diagonal basis:

\begin{equation}
\label{eq:offdiag}
\eta_{ij} = \eta^{ij} = \pmatrix{0&1&0\cr
                     1&0&0\cr
                     0&0&-1}\, .
\end{equation}

Note that the matrix $M^{-1}_{ij}$ itself is an $O(1,2)$ matrix since
it leaves invariant the metric $\eta$

\begin{equation}
(M^{-1})^{T} \eta M^{-1} = \eta\, .
\end{equation}

Under an $O(1,2)_{\rm Sugra}$ transformation $\Omega$ the vectors and
scalars transform in this way \cite{Ma1}:

\begin{equation}
\label{eq:trc}
{\cal A}_{\mu}^{\prime} = \Omega {\cal A}_{\mu}\, ,\hspace{1.2cm}
(M^{-1})^{\prime} = \Omega M^{-1} \Omega^{T}\, .
\end{equation}

Let us now consider the explicit form of the $O(1,2)_{\rm Sugra}$
transformations in more detail.  First of all, $O(1,2)_{\rm Sugra}=
SO^{\uparrow}(1,2)_{\rm Sugra}\times\ \Z^{(B)}_{2}\times\ \Z_{2}^{(S)}$.
The $\Z^{(B)}_{2}\times\ \Z^{(S)}_{2}$ transformations are again given
by Eqs.~(\ref{eq:dualreducedvector1},\ref{eq:dualreducedvector2}),
specified to the Abelian case.  The $\Z_{2}^{(B)}$ leads to the
generalized Buscher's rules given in Eqs.~(\ref{eq:dualvector}).  We
next consider the continuous $SO^{\uparrow}(1,2)_{\rm Sugra}$
transformations.  It is convenient to first consider the $so(1,2)$ Lie
algebra with generators $J_{1}, J_{2}$ and $J_{3}$:

\begin{equation}
{[}J_{1}, J_{2}{]} = J_{3}\, , \hspace{.3cm}
{[}J_{2}, J_{3}{]} = -J_{1}\, ,\hspace{.3cm}
{[}J_{3}, J_{1}{]} = J_{2}\, .
\end{equation}

\noindent In a $3\times 3$-matrix representation they satisfy the
symmetry property:

\begin{equation}
(J_{i}\eta)^{T} = - (J_{i}\eta)\, ,\hspace{1.2cm} i=1,2,3\, .
\end{equation}

\noindent In the off-diagonal basis Eq.~(\ref{eq:offdiag}) the
generators $J_{i}$ are represented by the following $3\times 3$
matrices\footnote{ Note that $J_{1}, J_{3}$ generate $so(1,1)$
subalgebras while $J_{2}$ generates an $so(2)$ subalgebra.}:

\begin{eqnarray}
J_{1} & = & {\textstyle\frac{1}{\sqrt{2}}}\pmatrix{0&0&-1\cr
                                                   0&0&1\cr
                                                   1&-1&0}\, ,
\hspace{1.5cm}
J_{2}  =  {\textstyle\frac{1}{\sqrt{2}}}\pmatrix{0&0&-1\cr
                                                   0&0&-1\cr
                                                  -1&-1&0\cr}\, ,
\nonumber \\
& &
\\
J_{3} & = & \pmatrix{1&0&0\cr
                     0&-1&0\cr
                     0&0&0}\, .
\nonumber
\end{eqnarray}

\noindent In terms of

\begin{equation}
J_{+} = (J_{2} + J_{1})/\sqrt{2}\, ,\hspace{1.2cm} J_{-} =
(J_{2}-J_{1})/\sqrt{2}\, ,
\end{equation}

\noindent we have the following commutation relations:

\begin{equation}
{[}J_{3}, J_{+}{]} = J_{+}\, ,\hspace{.3cm}
{[}J_{3}, J_{-}{]} = -J_{-}\, ,\hspace{.3cm}
{[}J_{+}, J_{-}{]} = J_{3}\, .
\end{equation}

The exponentiation of $J_{3},J_{+}$ and $J_{-}$ leads to the following
$SO^{\uparrow}(1,2)$ group elements:

\begin{eqnarray}
\exp{\alpha J_{3}} & = & \pmatrix{e^\alpha&0&0\cr
                                  0&e^{-\alpha}&0\cr
                                  0&0&1}\, ,
\nonumber \\
& &
\nonumber \\
\exp{\beta J_{-}} & = & \pmatrix{1&0&0\cr
                                 {1\over 2}\beta^2&1&-\beta\cr
                                 -\beta&0&1}\, ,
\\
& &
\nonumber \\
\exp{\gamma J_{+}} & = & \pmatrix{1&{1\over 2}\gamma^2&-\gamma\cr
                                  0&1&0\cr
                                  0&-\gamma&1}\, .
\nonumber
\end{eqnarray}

An arbitrary $SO^{\uparrow}(1,2)$ group element can be written as the
product of these basis elements.  Using Eqs.~(\ref{eq:trc}) one can
verify that the transformations in the basis above induce the following
transformations in $D-1$ dimensions.  First of all, the transformation
generated by $J_{3}$ in $D-1$ dimensions is just the
$SO^{\uparrow}(1,1)_{\rm Sugra}$ transformation of previous sections.
We next consider the transformation generated by $J_{-}$.  The
$(D-1)$-dimensional rules are given by

\begin{equation}
\begin{array}{rclrcl}
A_{\mu}^{\prime} & = & A_{\mu}\, ,&
(k^2)^{\prime} & = & k^2\, ,
\\
& &
\\
B_{\mu}^{\prime} & = & B_{\mu}- \beta V_{\mu}
+{\textstyle \frac{1}{2}}\beta^{2} A_{\mu}\, ,
\hspace{1.2cm}&
\ell^{\prime} & = & \ell+\beta\, ,
\\
& &
\\
V_{\mu}^{\prime} & = & V_{\mu} - \beta A_{\mu}\, . &
& &
\end{array}
\end{equation}

\noindent The corresponding transformation of the $D$--dimensional
fields is

\begin{eqnarray}
\hat{V}_{\underline{x}}^{\prime} & = &
\hat{V}_{\underline{x}} + \beta\, ,
\nonumber\\
& &
\nonumber\\
\hat{B}_{\underline{x}\mu}^{\prime} & = &
\hat{B}_{\underline{x}\mu} -{\textstyle \frac{1}{2}}\beta
\hat{V}_{\mu}\, .
\end{eqnarray}

\noindent All other fields are invariant.  It turns out that this
transformation is a special finite $U(1)$ gauge transformation
accompanied of a N--T transformation

\begin{eqnarray}
\hat{V}_{\hat\mu}^{\prime} & = &
\hat{V}_{\hat \mu} + \partial_{\hat\mu}\Lambda\, ,
\nonumber\\
& &
\nonumber\\
\hat{B}_{\hat\mu\hat\nu}^{\prime} & = &
\hat{B}_{\hat\mu\hat\nu} +\hat{V}_{[\hat\mu}
\partial_{\hat\nu]}\Lambda\, ,
\end{eqnarray}

\noindent with the parameter $\Lambda$ given by $\Lambda = \beta x$
\cite{Se1}.

Finally, we consider the transformation generated by $J_{+}$ . This is
the solution-generating transformation which was first introduced by
Sen \cite{Se1}.  The rules in $D-1$ dimensions are given by

\begin{equation}
\begin{array}{rclrcl}
A_{\mu}^{\prime} & = & A_{\mu} +{\textstyle \frac{1}{2}}\gamma^{2}
B_{\mu} -\gamma V_{\mu}\, ,
\hspace{1cm}&
(k^2)^{\prime} & = & \left (\frac{4k}{4 + 4\gamma \ell +
(\ell^{2}+2k^{2})\gamma^{2}}\right )^2\, ,
\\
& &
\\
B_\mu^\prime & = & B_\mu\, , &
\ell^{\prime} & = & \frac{4\ell +2 (\ell^{2}+2k^{2})\gamma}{4
+4\gamma \ell + (\ell^{2}+2k^{2})\gamma^{2}}\, ,
\\
& &
\\
V_{\mu}^{\prime} & = & -\gamma B_{\mu} + V_{\mu}\, . &
& &
\end{array}
\end{equation}

\noindent This transformation does not correspond to any gauge
transformation in $D$ dimensions. Indeed, as we have seen, from the
entire group $O(1,2)_{\rm Sugra}$, only this transformation and
Buscher's ($B$) are non-trivial solution generating transformations.
They correspond to the subgroup $O(1) \times O(2)$, while the other
transformations belong to the coset $O(1,2)_{\rm Sugra}/(O(1)\times O(2))$
of pure gauge transformations \cite{Se1}.

It is instructive to also consider the infinitesimal form of the
$SO^{\uparrow}(1,2)$ transformations in $D-1$ dimensions:

\begin{eqnarray}
\delta A_\mu & = & \alpha A_\mu - \gamma V_\mu\, ,
\nonumber\\
&&
\nonumber\\
\delta B_\mu & = & -\alpha B_\mu - \beta V_\mu\, ,
\nonumber\\
&&
\nonumber\\
\delta V_\mu & = & -\beta A_\mu -\gamma B_\mu\, ,
\nonumber\\
&&
\nonumber\\
\delta k^2 & = & -2\alpha k^2 - 2\gamma \ell k^2\, ,
\nonumber\\
&&
\nonumber\\
\delta \ell & = & -\alpha \ell + \beta - {\textstyle \frac{1}{2}}
\gamma \bigl (\ell^2 - 2k^2\bigr )\, .
\end{eqnarray}

\noindent These infinitesimal rules lead to the commutation relations of
the $so(1,2)$ algebra:

\begin{eqnarray}
{[}\delta_{\alpha}, \delta_{\beta} {]} & = &
\delta_{\beta^{\prime}}
\hspace{1.2cm} {\rm with}\ \ \ \beta^{\prime}=\alpha\beta\, ,
\nonumber\\
& &
\nonumber\\
{[}\delta_{\alpha}, \delta_{\gamma} {]} & = &
\delta_{\gamma^{\prime}}
\hspace{1.2cm} {\rm with}\ \ \ \gamma^{\prime}=-\alpha\gamma\, ,
\\
& &
\nonumber \\
{[}\delta_{\beta}, \delta_{\gamma} {]} & = &
\delta_{\alpha^{\prime}}
\hspace{1.2cm} {\rm with}\ \ \ \alpha^{\prime}=\gamma\beta\, .
\nonumber
\label{eq:commrel}
\end{eqnarray}

\noindent One may verify that the action (\ref{eq:actionvectorreduced})
is indeed left invariant by the infinitesimal transformations given
above.

The rules of transformation of the scalars $k^{2}$ and $\ell$ are
complicated and they lead to even more complicated rules for the
$D$--dimensional fields that are given in Appendix A. They considerably
simplify if we assume that $\hat{V}_{\mu} = \hat{V}_{\underline{x}} =
0$\footnote{The $J_{+}$--transformation is often used to construct a
charged solution $(\hat{V}^\prime \ne 0)$ out of an uncharged one
$(\hat{V} = 0)$.  Therefore, most times it is enough to know the
$J_{+}$--transformation for the case that $\hat{V}_{\mu}
=\hat{V}_{\underline{x}} = 0$.}.  In that case they are given by

\begin{equation}
\begin{array}{rclrcl}
\hat{g}_{\mu\nu}^{\prime} & = &
\hat{g}_{\mu\nu} -\frac{\hat{g}_{\underline{x}\mu}
\hat{g}_{\underline{x}\nu}}{\hat{g}_{\underline{xx}}}
+\frac{4 \hat{g}_{\underline{xx}}}{(2
-\gamma^{2} \hat{g}_{\underline{xx}})^{2}}
\left(\frac{\hat{g}_{\underline{x}\mu}}{\hat{g}_{\underline{xx}}}
+{\textstyle \frac{1}{2}}\gamma^{2}\hat{B}_{\underline{x}\mu}\right)
\left(\frac{\hat{g}_{\underline{x}\nu}}{\hat{g}_{\underline{xx}}}
+{\textstyle\frac{1}{2}}\gamma^{2}\hat{B}_{\underline{x}\nu}\right)\, ,
\!\!\!\!\!\!\!\!\!\!\!\!\!\!\!\!\!\!\!\!\!\!\!\!\!\!\!\!\!\!\!\!\!\!\!
\!\!\!\!\!\!\!\!\!\!\!\!\!\!\!\!\!\!\!\!\!\!\!\!\!\!\!\!\!\!\!\!\!\!\!
\!\!\!\!\!\!\!\!\!\!\!\!
& & &
\\
& & & & &
\\
\hat{B}_{\mu\nu}^{\prime} & = & \hat{B}_{\mu\nu}+2\gamma^{2}
\frac{\hat{g}_{\underline{x}[\mu}\hat{B}_{\underline{x}\nu]}}{
2 -\gamma^{2}{\hat g}_{\underline{xx}}}\, ,&
\hat{B}_{\underline{x}\mu}^{\prime} & = &
\frac{2 {\hat B}_{{\underline x}\mu}}{2 -
\gamma^2 {\hat g}_{{\underline {xx}}}}\, ,
\\
& & & & &
\\
\hat{g}_{\underline{x}\mu}^{\prime} & = &
\frac{4\hat{g}_{\underline{xx}}}{(2
-\gamma^{2} \hat{g}_{\underline{xx}})^{2}}
\left( \frac{\hat{g}_{\underline{x}\mu}}{\hat{g}_{\underline{xx}}}
+{\textstyle\frac{1}{2}}\gamma^{2}
\hat{B}_{\underline{x}\mu}\right)\, , &
\hat{g}_{\underline{xx}}^{\prime} & = &
\frac{4 \hat{g}_{\underline{xx}}}{(2
-\gamma^{2}\hat{g}_{\underline{xx}})^{2}}\, ,
\\
& & & & &
\\
\hat{V}_{\mu}^{\prime} & = &
-\gamma\hat{B}_{\underline{x}\mu}
-2\gamma \frac{\hat g_{\underline{xx}}}{2 -\gamma^{2}
{\hat g}_{\underline{xx}}}
\left( \frac{\hat{g}_{\underline{x}\mu}}{\hat{g}_{\underline{xx}}}
+{\textstyle \frac{1}{2}}\gamma^{2}\hat{B}_{\underline{x} \mu}
\right)\, ,
\!\!\!\!\!\!\!\!\!\!\!\!\!\!\!\!\!\!\!\!\!\!\!\!\!\!\!\!\!\!\!\!\!\!\!
& & &
\\
& & & & &
\\
\hat{V}_{\underline{x}}^{\prime} & = &
-2\gamma \frac{\hat g_{\underline{xx}}}{2
-\gamma^{2}{\hat g}_{\underline{xx}}}\, , &
\hat{\phi}^{\prime} & = &
\hat{\phi} +{\textstyle\frac{1}{2}}
\log \frac{2}{2-\gamma^2{\hat g}_{\underline{xx}}}\, .
\end{array}
\label{eq:hor}
\end{equation}

\noindent The above formulae have been given in Eq.~(2.12) of
Ref.~\cite{Ho1} for the special case that $\hat{g}_{\underline{x}\mu} =
\hat{B}_{\underline{x}\mu} = 0$.  In that case the fields
$\hat{g}_{\underline{x}\mu}, \hat{B}_{\underline{x}\mu}$ and
$\hat{V}_{\mu}$ remain zero after the Sen transformation;
$\hat{g}_{\mu\nu}$ and $\hat{B}_{\mu\nu}$ remain invariant while the
other fields transform as follows:

\begin{eqnarray}
\hat{g}_{\underline{xx}}^{\prime} & = &
\frac{4\hat{g}_{\underline{xx}}}{(2
-\gamma^{2}\hat{g}_{\underline{xx}})^{2}}\, ,
\nonumber \\
& &
\nonumber \\
\hat{V}_{\underline{x}}^{\prime} & = &
-2\gamma \frac{\hat{g}_{\underline{xx}}}{2
-\gamma^{2}{\hat g}_{\underline{xx}}}\, ,
\\
& &
\nonumber \\
\hat{\phi}^{\prime} & = &
\hat{\phi} +{\textstyle \frac{1}{2}}\log
\frac{2}{2-\gamma^{2}{\hat g}_{\underline{xx}}}\, .
\nonumber
\end{eqnarray}

\noindent These formulae differ from Eq.~(2.12) of \cite{Ho1}.  To obtain
the formula of Ref.~\cite{Ho1} (with parameter $\alpha_{H}$) one should
perform successively an $\alpha$, $\beta$ and $\gamma$
trans\-for\-ma\-tion with parameters\footnote{Note that the
normalization of the vector field $V_\mu$ in \cite{Ho1} differs from
ours with a factor of $2i$.}

\begin{equation}
\alpha = 2{\rm log}\ {\rm cosh}\ \alpha_H\, \hspace {.5cm}
\beta = -i \sqrt 2 \frac{{\rm sinh}\ \alpha_H}{{\rm cosh}^3\
\alpha_H}\, \hspace {.5cm}
\gamma = i\sqrt 2 {\rm tanh}\ \alpha_H\, .
\end{equation}

Finally, the infinitesimal form of the $SO^{\uparrow}(1,2)$
tranformations in $D$ dimensions is given by:

\begin{eqnarray}
\label{eq:inf}
\delta \hat{g}_{\underline{xx}} & = &
-2\alpha\hat{g}_{\underline{xx}}
-2\gamma\hat{V}_{\underline{x}}\hat{g}_{\underline{xx}}\, ,
\nonumber\\
& &
\nonumber\\
\delta \hat{g}_{\underline{x}\mu} & = &
-\alpha \hat{g}_{\underline{x}\mu}
-\gamma \left( \hat{V}_\mu\hat{g}_{\underline{xx}}
+ \hat{V}_{\underline{x}} \hat{g}_{\underline{x}\mu}\right)\, ,
\nonumber\\
& &
\nonumber\\
\delta \hat{g}_{\mu\nu} & = &
-2\gamma\hat{V}_{(\mu}\hat{g}_{\underline{x}\nu)}\, ,
\nonumber\\
& &
\nonumber\\
\delta \hat{B}_{\underline{x}\mu} & = &
-\alpha \hat{B}_{\underline{x}\mu}
-{\textstyle \frac{1}{2}}\beta \hat{V}_{\mu}
-{\textstyle\frac{1}{2}}\gamma\left(
\hat{V}_{\underline{x}}\hat{B}_{\underline{x}\mu}
-\hat{g}_{\underline{x}\mu}\hat{V}_{\underline{x}}
+\hat{g}_{\underline{xx}} \hat{V}_{\mu}\right)\, ,
\nonumber\\
& &
\nonumber\\
\delta \hat{B}_{\mu\nu} & = &
-\gamma\left(\hat{V}_{[\mu}\hat{B}_{\underline{x}\nu]}
+\hat{g}_{\underline{x}[\mu}\hat{V}_{\nu]}\right)\, ,
\\
& &
\nonumber \\
\delta \hat{V}_{\underline{x}} & = &
-\alpha \hat{V}_{\underline{x}}
+\beta -\gamma \left( g_{\underline{xx}}
+{\textstyle \frac{1}{2}}\hat{V}_{\underline{x}}^{2} \right)\, ,
\nonumber\\
& &
\nonumber\\
\delta \hat{V}_{\mu} & = &
-\gamma\left(\hat g_{\underline{x}\mu} +\hat{B}_{\underline{x}\mu}
+{\textstyle \frac{1}{2}}
\hat{V}_{\underline{x}}\hat{V}_{\mu}\right)\, ,
\nonumber\\
& &
\nonumber\\
\delta \hat{\phi} & = &
-{\textstyle\frac{1}{2}}\alpha -{\textstyle\frac{1}{2}}\gamma
\hat{V}_{\underline{x}}\, .
\nonumber
\end{eqnarray}

\noindent As in the $(D-1)$-dimensional case (see above) these
infinitesimal rules lead to the commutation relations of the $so(1,2)$
algebra given in (\ref{eq:commrel}).

\noindent To conclude we give the full symmetry group of the equations
of motion:

\begin{equation}
SO^{\uparrow}(1,2)_{\rm Sugra} \times\ SO^{\uparrow}(1,1)_{x-y} \times\
SO{\uparrow}(1,1)_{\rm string} \times\ \Z_2^{(B)} \times\ \Z_2^{(S)}.
\end{equation}


\section{$\alpha^{\prime}$ Corrections}
\label{sec-alpha}

In the first section we considered the zero-slope limit of the bosonic
sector he\-te\-ro\-tic string effective action and in the following
sections we have added Yang--Mills fields to it.  This is consistent
from the supergravity point of view.  However, from the heterotic string
theory point of view, the Yang--Mills term is already first order in
$\alpha^{\prime}$ and, strictly speaking, one has to add to the action
all the other terms linear in $\alpha^{\prime}$.  Therefore, to this
order, the action that we have to consider is \cite{Be1}

\begin{eqnarray}
S^{(D)}_{{\rm Sugra}+\alpha'} & = & {\textstyle\frac{1}{2}}\int d^{D}x\
\sqrt{-\hat{g}}\ e^{-2\hat{\phi}}\left\{
-\hat{R}+4(\partial\hat{\phi})^{2}-
{\textstyle\frac{3}{4}}(\hat{H}^{(1)})^{2}\right.
\nonumber \\
& &
\nonumber \\
& &
\left.+\ {\textstyle\frac{1}{4}} \left[
\beta{\rm Tr} \ \hat{F}^{2}\left(\hat V\right) +
\alpha {\rm Tr} \ \hat{R}^{2}\left( \hat{\Omega}^{(0)}\right)
\right]\right\}\, ,
\label{eq:s1}
\end{eqnarray}

\noindent Here $\hat{H}^{(1)}$ is the axion field strength up to linear
order in $\alpha^\prime$

\begin{equation}
\hat{H}^{(1)} = \hat{H}^{(0)} -\left(\beta\hat{\omega}^{YM}
+\alpha\hat{\omega}^{(0)L}\right)\, ,
\end{equation}

\noindent $\hat{H}^{(0)}$ is the zero order in $\alpha^{\prime}$ axion
field strength, $\hat\omega^{YM}$ is the Yang--Mills Chern--Simons form
and $\hat{\omega}^{(0)L}$ is the (zero order in $\alpha^{\prime}$)
Lorentz Chern--Simons form.  They are respectively given by\footnote{We
use a short-hand notation in which the antisymmetrized world indices are
not indicated.}

\begin{eqnarray}
\label{eq:ol}
\hat{H}^{(0)} & = & \partial\hat{B}\, ,
\nonumber \\
& &
\nonumber \\
\hat \omega^{YM} & = &
{\textstyle\frac{1}{2}}\hat{V}^{I}\hat{F}_{I}
\left(\hat{V} \right) + {\textstyle\frac{1}{6}} f_{IJK} \hat{V}^{I}
\hat{V}^{J} \hat{V}^{K}\, ,
\\
& &
\nonumber \\
\hat{\omega}^{(0)L} & = &
{\textstyle\frac{1}{2}} \hat{\Omega}^{(0)\hat{a}\hat{b}}
\hat{R}^{\hat{a}\hat{b}}\left( \hat{\Omega}^{(0)} \right)
+{\textstyle\frac{1}{3}} \hat{\Omega}^{(0)\hat{a}\hat{b}}\
\hat{\Omega}^{(0)\hat{a}\hat{c}}\ \hat{\Omega}^{(0)\hat{c}\hat{b}}\, .
\nonumber
\end{eqnarray}

\noindent where the (zero order in $\alpha^{\prime}$) torsionful spin
connection $\hat{\Omega}^{(0)}$ is defined by

\begin{equation}
\hat{\Omega}_{\hat\mu}{}^{(0)\hat{a}\hat{b}} =
\hat\omega_{\hat\mu}{}^{\hat{a}\hat{b}}
+{\textstyle\frac{3}{2}}\hat{H}_{\hat{\mu}}{}^{(0)\hat{a}\hat{b}}\, .
\end{equation}

\noindent Finally, $\alpha$ and $\beta$ are constants which are related
to $\alpha^\prime$ as follows

\begin{equation}
\alpha = 2\alpha^{\prime}\, ,\hspace{1.5cm}
\beta = {\textstyle\frac{1}{15}} \alpha^{\prime}\, .
\end{equation}

\noindent Note that the action used in the previous sections can be
obtained by setting $\alpha = 0$ and $\beta = 1/g^{2}$.

Once we assume that all the solutions to the equations of motion derived
from the above action have an isometry, we expect the usual duality
group $SO(1,1)_{\rm Sugra}$.  The question we really need to address now
is whether a generalization of Buscher's duality transformation exists.

This transformation should coincide with the $B$ transformations found
in previous sections in the appropriate limits (Buscher's original
transformations Eqs.~(\ref{eq:Buscher}) in the limit
$\alpha^{\prime}\rightarrow 0$ and the generalization
Eqs.~(\ref{eq:dualvector},\ref{eq:Ghat})).  Then, we are actually
looking for the complete form of the $\alpha^{\prime}$ corrections to
Eqs.~(\ref{eq:Buscher}) and we know the contribution of the vector
fields to them.

We are now going to argue, using a simple observation, that the
corrections to the $T$~duality rules can be obtained in a
straightforward manner.  We first note that the torsionful spin
connection $\hat{\Omega}^{(0)}$ is a dependent field given in terms of
the $D$--bein and the axion.  This fixes the zero-th order duality rules
of $\hat {\Omega}^{(0)}$.  To calculate these we first have to give the
duality rules of the $D$--bein.  For this purpose we parametrize the
$D$--bein as follows:

\begin{equation}
(\hat{e}_{\hat{\mu}}{}^{\hat{a}})=
\left(
\begin{array}{cc}
e_{\mu}{}^{a} & k A_{\mu} \\
0           & k
\end{array}
\right)
\, ,
\hspace{1cm}
(\hat{e}_{\hat{a}}{}^{\hat{\mu}})=
\left(
\begin{array}{cc}
e_{a}{}^{\mu} & -A_{a} \\
0           & k^{-1}
\end{array}
\right)\, ,
\label{eq:basis}
\end{equation}

\noindent where $A_{a}=e_{a}{}^{\mu}A_{\mu}$.  Note that this is the
first time that we are forced to use $k$ instead of $k^2$.  To lowest
order in $\alpha^\prime$ the duality rule of $k^2$ is given by
$\tilde{k^2} = 1/k^2$.  This means that for $k$ we have

\begin{equation}
\tilde k = \mp \frac{1}{k}\, .
\label{eq:dualk}
\end{equation}

These two signs are not really different since the two possibilities are
related to each other by a discrete Lorentz transformation (in tangent
space) $\tilde{\hat{e}}_{\hat{\mu}}{}^{x} =-\hat{e}_{\hat{\mu}}{}^{x}$.
In $D$ dimensions this leads to the following lowest-order rule of the
$D$-bein, (here we choose the upper sign in Eq.~(\ref{eq:dualk})):

\begin{eqnarray}
\tilde {\hat e}_{\underline x}{}^{\hat a} &=&
\frac{1}{\hat g_{\underline{xx}}}
\hat e_{\underline x}{}^{\hat a}\, ,\nonumber\\
\tilde {\hat e}_\mu{}^{\hat a} &=& \hat e_\mu{}^{\hat a}
-\frac{1}{\hat g_{\underline{xx}}}
\biggl ( \hat g_{\underline x\mu} - \hat B_{\underline x\mu}\biggr )
\hat e_{\underline x}{}^{\hat a}\, ,
\label{eq:dualDbein}
\end{eqnarray}

\noindent so the dual of $\hat \Omega^{(0)}$ is

\begin{equation}
 \tilde{\hat{\Omega}}_{\underline{x}}{}^{(0)\hat{a}\hat{b}} =
-\hat{\Omega}_{\underline{x}}{}^{(0)\hat{a}\hat{b}} /
\hat{G}_{\underline{xx}}^{(0)}\, ,
\hspace{.5cm}
\tilde{\hat{\Omega}}_{\mu}{}^{(0)\hat{a}\hat{b}} =
\hat{\Omega}_{\mu}{}^{(0)\hat{a}\hat{b}}
-\hat{\Omega}_{\underline{x}}{}^{(0)\hat{a}\hat{b}}
\hat{G}_{\underline{x}\mu}^{(0)}/\hat{G}^{(0)}_{\underline{xx}}\, ,
\end{equation}

\noindent where $\hat{G}^{(0)}_{\hat{\mu}\hat{\nu}}
=\hat{g}_{\hat\mu\hat\nu} +\hat{B}_{\hat\mu\hat\nu}$ is the zero-slope
limit of $\hat{G}_{\hat\mu\hat\nu}$.  We now observe that this duality
rule is identical to that of a non-Abelian vector field $\hat V_{\hat
\mu}^I$ (see Eq.~(\ref{eq:dualvector})), to lowest order in
$\alpha^\prime$, when we consider the pair of Lorentz indices
$\hat{a}\hat{b}$ as a Yang--Mills index.

\noindent We can combine this with the observation that we already know
how to construct duality-invariant actions for the Yang--Mills fields.
In fact we can extend our results for the Yang--Mills fields to a more
general action formula: given a vector field which, to lowest order in
$\alpha^\prime$, transforms as given in Eqs.~(\ref{eq:dualvector}), an
action can be constructed which is duality invariant up to linear order
in $\alpha^\prime$.  The action is given by Eq.~(\ref{eq:actionvector})
(with the identification $1/g^2 = \beta = \alpha^{\prime}/15$ and the
Yang--Mills field being replaced by the vector field in question) and
the corresponding duality rules are given by Eqs.~(\ref{eq:dualvector}).

We now apply the above action formula to the case that the gauge group
of the vector field is given by the direct product of the Yang--Mills
group times the $D$-dimensional Lorentz group.  This leads to the action
given in Eq.~(\ref{eq:s1}).  The corresponding duality rules, to linear
order in $\alpha^\prime$ are given by Eqs.~(\ref{eq:dualvector}), where

\begin{equation}
\hat{G}_{\hat{\mu}\hat{\nu}}=
\hat{g}_{\hat{\mu}\hat{\nu}}+\hat{B}_{\hat{\mu}\hat{\nu}}
-{\textstyle\frac{1}{2}}
\left\{
\alpha\hat{\Omega}_{\hat{\mu}}{}^{(0)\hat{a}\hat{b}}
\hat{\Omega}_{\hat{\nu}}{}^{(0)\hat{a}\hat{b}}
+\beta\hat{V}^{I}_{\hat{\mu}}\hat{V}_{I\hat{\nu}}
\right\}\, .
\label{eq:Ghatnonabeliancomplete}
\end{equation}

\noindent instead of Eq.~(\ref{eq:Ghat}).

We would like to stress the following points:

\begin{itemize}

\item The duality rules that we just have obtained considerably simplify
if the gauge group is embedded into the holonomy group since in that
case the last two terms in Eq.~(\ref{eq:Ghatnonabeliancomplete}) cancel
against each other.  We note that for $\alpha = 0$ and Abelian vector
fields, the duality rules of the gauge fields are those of
Refs.~\cite{Sha1,Gi2}.  These rules can be derived using the
$\sigma$-model approach if the gauge fields couple to the string via
bosonic group coordinates.  The same rules can also be derived for the
case that the gauge fields couple via heterotic fermions to the string.
However, in that case, to obtain the same answer, one has to take into
account the Yang--Mills anomaly.  In the general case with $\alpha \ne
0$ one also should consider the Lorentz anomaly.  In case the embedding
is made, there is an anomaly cancellation which leads to the simplified
duality rules mentioned above.  In particular, the simplified duality
rules of the vector fields are now the ones given in Ref.~\cite{Be4}.

\item Since the structure of the higher order in $\alpha^{\prime}$
corrections seems to be such that the torsionful spin connection
$\hat{\Omega}$ enters always in the same way as the Yang-Mills vector
field $\hat{V}^{I}$ (apart from the fact that $\hat{\Omega}$ has to be
redefined at each order but $\hat{V}^{I}$ does not), one may expect that
the structure of higher order in $\alpha^{\prime}$ corrections to the
duality rules will be such that Eqs.~(\ref{eq:dualvector}) can still be
used but the effective metric (\ref{eq:Ghatnonabeliancomplete}) will get
higher order corrections in which $\hat{\Omega}$ and $\hat{V}^{I}$ will
appear in the same way.  If this was true, the embedding of the gauge
group into the holonomy group would produce a cancellation of all the
corrections and Buscher's original transformations would not get
corrections.  This is also consistent with the results in
Ref.~\cite{Be4}.

\item It is interesting to note that a combination similar to the
effective metric given in Eq.~(\ref{eq:Ghatnonabeliancomplete}) also
appears in Ref.~\cite{Hu3}.  There it was observed that a manifestly
supersymmetric way of cancelling the Green-Schwarz anomaly in the
heterotic string effective action is to make a redefinition of the
metric.  The new metric is essentially our effective metric. This
suggests that this effective metric could play an important role in the
heterotic string effective action and that it could be the right object
in terms of which many terms could be expressed. A geometrical or
physical interpretation is still lacking.

\end{itemize}

We thus conclude that Buscher's duality transformations have a
straightforward generalization to first order in $\alpha^{\prime}$.  Are
the other zero-slope duality symmetries also preserved?  The answer is
yes (except for the $A$ transformation).  The duality symmetries are
then those of the Sugra+YM action.

Finally, we can also ask what happens to the $SO(1,2)_{\rm Sugra}$
duality group if we take just a single Abelian field.  Do these
transformations receive $\alpha^{\prime}$ corrections as well or are
they already exact up to this order?  The situation is not entirely
clear: it is true that both the $\alpha$ as the $\beta$ transformations
(the ones that are gauge transformations and g.c.t.'s) do not receive
corrections and we suppose that the $so(1,2)$ algebra holds.  However
this does not mean that the non-trivial solution generating
transformation $\gamma$ cannot have corrections because of the algebra
structure.  It would be entirely consistent with the absence of
$\alpha^{\prime}$ corrections in the $\alpha,\beta$ transformations and
in the $so(1,2)$ algebra to assume that the $\gamma$ transformations
have $\alpha^{\prime}$ corrections that commute with the $\alpha$ and
$\beta$ transformations.  More work is necessary to answer these
questions and we hope to present our results elsewhere soon \cite{Ja1}.


\section*{Acknowledgements}

We are grateful to Chris Hull and Arkadi Tseytlin for most fruitful
discussions.  One of us (T.O.)~is extremely grateful to the hospitality,
friendly environment and financial support of the Institute for
Theoretical Physics of the University of Groningen and the Physics
Department of Stanford University, where part of this work was done.
The work of E.B.~has been made possible by a fellowship of the Royal
Netherlands Academy of Arts and Sciences (KNAW).  He also thanks the
Physics department of Stanford University for hospitality.  The work of
E.B., and T.O.~has also been partially supported by a NATO Collaboration
Research Grant.  The work of T.O. was supported by a European Union {\it
Human Capital and Mobility} program grant.  The work of B.J.~was
performed as part of the research program of the ``Stichting voor
Fundamenteel Onderzoek der Materie'' (FOM).


\appendix


\section{The Sen Transformations}
\label{sec-ST}

In this appendix we give the explicit form of the solution generating
transformation introduced by Sen in $D$ dimensions.  A special case plus
the infinitesimal form of these formulae are given in
Eqs.~(\ref{eq:hor}) and Eqs.~(\ref{eq:inf}), respectively.  The general
and finite Sen rules are:

\begin{eqnarray}
\hat{g}_{\underline{xx}}^{\prime} & = &
{\textstyle\frac{16}{N^{2}}}\hat{g}_{\underline{xx}}\, ,
\nonumber\\
& &
\nonumber\\
\hat{g}_{\underline{x}\mu}^{\prime}
& = &
{\textstyle\frac{16}{N^{2}}}\hat{g}_{\underline{xx}}
\left[
\frac{\hat{g}_{\underline{x}\mu}}{\hat{g}_{\underline{xx}}}
-\gamma
\left(
\hat{V}_{\mu} -\hat{V}_{\underline{x}}
\frac{\hat{g}_{\underline{x}\mu}}{\hat{g}_{\underline{xx}}}
\right)
\right.
\nonumber\\
& &
\nonumber\\
& &
\left.
+{\textstyle\frac{1}{2}}\gamma^{2}
\left(
\hat{B}_{\underline{x}\mu}
-{\textstyle\frac{1}{2}}\hat{V}_{\underline{x}}\hat{V}_{\mu}
+{\textstyle\frac{1}{2}}\hat{V}_{\underline{x}}^{2}
\frac{\hat{g}_{\underline{x}\mu}}{\hat{g}_{\underline{xx}}}
\right)
\right]\, ,
\nonumber \\
& &
\nonumber\\
\hat{g}_{\mu\nu}^{\prime} & = &
\hat{g}_{\mu\nu} -\frac{\hat{g}_{\underline{x}\mu}
\hat{g}_{\underline{x}\nu}}{\hat{g}_{\underline{xx}}}
\nonumber\\
& &
\nonumber\\
& &
\hspace{-1cm}
+{\textstyle\frac{16}{N^{2}}}\hat{g}_{\underline{xx}}
\left[
\frac{\hat{g}_{\underline{x}\mu}}{\hat{g}_{\underline{xx}}}
-\gamma
\left(
\hat{V}_{\mu} -\hat{V}_{\underline{x}}
\frac{\hat{g}_{\underline{x}\mu}}{\hat{g}_{\underline{xx}}}
\right)
+{\textstyle\frac{1}{2}}\gamma^{2}
\left(
\hat{B}_{\underline{x}\mu} -{\textstyle\frac{1}{2}}
\hat{V}_{\underline{x}}\hat{V}_{\mu}
+{\textstyle\frac{1}{2}}\hat{V}_{\underline{x}}^{2}
\frac{\hat{g}_{\underline{x}\mu}}{
\hat{g}_{\underline{xx}}}
\right)
\right]
\nonumber\\
& &
\nonumber\\
& &
\times
\left[
\frac{\hat{g}_{\underline{x}\nu}}{\hat{g}_{\underline{xx}}}
-\gamma
\left(
\hat{V}_\nu -\hat{V}_{\underline{x}}
\frac{\hat{g}_{\underline{x}\nu}}{\hat{g}_{\underline{xx}}}
\right)
+{\textstyle\frac{1}{2}}\gamma^{2}
\left(
\hat{B}_{\underline{x}\nu}-{\textstyle\frac{1}{2}}
\hat{V}_{\underline{x}} \hat{V}_{\nu}
+{\textstyle\frac{1}{2}}\hat{V}_{\underline{x}}^{2}
\frac{\hat{g}_{\underline{x}\nu}}{\hat{g}_{\underline{xx}}}
\right)
\right]\, ,
\nonumber\\
& &
\nonumber\\
\hat{B}_{\underline{x}\mu}^{\prime}
& = &
\hat{B}_{\underline{x}\mu} -{\textstyle\frac{1}{2}}\hat{V}_{\underline
x}\hat{V}_\mu + {\textstyle\frac{1}{2}}\hat V_{\underline{x}}^{2}
{\hat{g}_{\underline{x}\mu}\over \hat g_{\underline{xx}}}
+{\textstyle\frac{1}{N}}
\left[
2\hat{V}_{\underline{x}}+\gamma
\left(
\hat{V}_{\underline x}^{2}-2\hat{g}_{\underline{xx}})
\right)
\right]
\nonumber\\
& &
\nonumber\\
& &
\times
\left[
\hat{V}_\mu - \hat{V}_{\underline {x}}
{\hat{g}_{\underline{x}\mu}\over \hat{g}_{\underline{xx}}}
-\gamma
\left(
\hat B_{\underline{x}\mu}-{\textstyle\frac{1}{2}}
\hat{V}_{\underline{x}}
\hat{V}_\mu + {\textstyle\frac{1}{2}}\hat{V}_{\underline{x}}^2
{\hat{g}_{\underline{x}\mu}\over \hat{g}_{\underline{xx}}}
\right)
\right]\, ,
\nonumber\\
& &
\nonumber\\
\hat B_{\mu\nu}^{\prime}
& = &
\hat{B}_{\mu\nu} - \hat{V}_{\underline{x}}
{\hat{g}_{\underline{x}[\mu}\hat{V}_{\nu]}\over \hat{g}_{\underline{xx}}}
- \gamma
\left(
\hat{V}_{[\mu} -\hat{V}_{\underline {x}}{\hat{g}_{\underline{x}[\mu}
\over \hat{g}_{\underline{xx}}}
\right)
\hat B_{\underline{x}\nu]}
\nonumber \\
& &
\nonumber\\
& &
+{\textstyle\frac{2}{N}}
\left[
2\hat{V}_{\underline{x}}  +\gamma
\left(
\hat{V}_{\underline{x}}^{2} -2\hat{g}_{\underline{xx}}
\right)
\right]
\nonumber\\
& &
\nonumber\\
& &
\times
\left[
{\hat{g}_{\underline{x}[\mu}\hat{V}_{\nu]}\over
\hat{g}_{\underline{xx}}} -\gamma{\hat{g}_{\underline{x}[\mu}\over
\hat{g}_{\underline{xx}}}
\left(
\hat{B}_{\underline{x}\nu]}-{\textstyle\frac{1}{2}}
\hat{V}_{\underline{x}}\hat{V}_{\nu]}
\right)
+{\textstyle\frac{1}{2}}\gamma^{2}
\left(
\hat{V}_{[\mu}-\hat{V}_{\underline{x}}
{\hat{g}_{\underline {x}[\mu}\over \hat{g}_{\underline{xx}}}
\right)
\hat{B}_{\underline{x}\nu]}
\right]\, ,
\nonumber\\
& &
\nonumber\\
\hat{V}_{\underline{x}}^{\prime}
& = &
{\textstyle\frac{2}{N}}
\left[
2\hat{V}_{\underline x}+\gamma
\left(
\hat{V}_{\underline{x}}^{2}- 2\hat{g}_{\underline{xx}}
\right)
\right]\, ,
\nonumber\\
& &
\nonumber\\
\hat{V}_{\mu}^{\prime}
& = &
\hat{V}_{\mu} - \hat{V}_{\underline{x}}
{\hat{g}_{\underline{x}
\mu}\over \hat{g}_{\underline{xx}}} - \gamma
\biggl (\hat B_{\underline x \mu} -
{\textstyle\frac{1}{2}}\hat{V}_{\underline x}\hat{V}_{\mu}
+{\textstyle\frac{1}{2}}\hat V^2_{\underline {x}}
{\hat{g}_{\underline x \mu}\over \hat{g}_{\underline{xx}}}\biggr )
\nonumber\\
& &
\nonumber\\
& &
+{\textstyle\frac{2}{N}}
\left[
2\hat{V}_{\underline {x}}+\gamma
\left(
\hat{V}_{\underline x}^2- 2\hat{g}_{\underline{xx}}
\right)
\right]
\nonumber\\
& &
\nonumber\\
& &
\times
\left[
{\hat{g}_{\underline{x}\mu}\over \hat{g}_{\underline{xx}}}
- \gamma
\left(
\hat{V}_\mu -\hat{V}_{\underline{x}}{\hat{g}_{\underline {x}
\mu}\over \hat{g}_{\underline{xx}}}
\right)
+{\textstyle\frac{1}{2}}\gamma^{2}
\left(
\hat B_{\underline{x} \mu} -
{\textstyle\frac{1}{2}}\hat{V}_{\underline x}\hat{V}_\mu
+{\textstyle\frac{1}{2}}\hat{V}^2_{\underline {x}}
{\hat{g}_{\underline{x} \mu}\over \hat{g}_{\underline{xx}}}
\right)
\right]\, ,
\nonumber \\
& &
\nonumber\\
\hat{\phi}^{\prime}
&=&
\hat{\phi}+{\textstyle\frac{1}{2}}\log{\textstyle\frac{4}{N}}
\nonumber\, .
\end{eqnarray}

\noindent where

\begin{equation}
N = 4 + 4\gamma \hat{V}_{\underline x} + \gamma^2
\left(
\hat{V}_{\underline x}^2 -2 \hat{g}_{\underline{xx}}
\right)\, .
\end{equation}

Note that, unlike the case of the Buscher transformations, the effective
metric $\hat{G}_{\hat\mu\hat\nu}$ defined in (\ref{eq:Ghat}) does not
seem to play any special role in the above transformations.  We have
verified that under $SO(1,2)$ the effective metric does not transform
into itself.  This is in contradistinction with the $\Z_2$
transformations (see Eq.~(\ref{eq:simple})).


\section{Duality Symmetries In $D=11,10$ And $9$ Type~II Theories}
\label{sec-SA}

In this Appendix we will discuss duality symmetries in eleven, ten
and nine dimensions for Type~II theories. We will use the results and
conventions of Ref.~\cite{Be3}.

This Appendix is organized in subsections with increasing number of
isometries and decreasing number of dimensions for each case.

\subsection{No isometries}

\begin{description}

\item[D=11] There is a single $SO^{\uparrow}(1,1)_{\rm brane}$ symmetry
whose weights are given in Table~\ref{tab:d11zero}.  This symmetry
essentially counts the mass dimension of the different fields.  We also
stress that $\hat{\hat{C}}$ is a pseudo-tensor that changes sign under
improper g.c.t.~'s.

\begin{table}
\begin{center}
\begin{tabular}{||c||c|c|c||}
\hline\hline
& & & \\
Name & $\hat{\hat{C}}$ & $\hat{\hat{g}}$ & $S^{(11)}$ \\
\hline\hline
& & & \\
$SO^{\uparrow}(1,1)_{\rm brane}$ & $\frac{3}{2}$ & $1$ & $\frac{9}{2}$ \\
& & & \\
\hline\hline
\end{tabular}
\end{center}
\caption{\textit{Weights of the $D=11,N=1$ supergravity fields and
action under $SO^{\uparrow}(1,1)_{\rm brane}$.}}
\label{tab:d11zero}
\end{table}

\end{description}

\subsection{One isometry}

\begin{description}

\item[D=11] In addition to the symmetries of the previous section, we
have to consider the subgroup of g.c.t.~'s that preserve the condition
that the fields do not depend on the coordinate $y$. This group is

\begin{equation}
GL(1,\R)=SO^{\uparrow}(1,1)\times\ \Z^{(y)}_{2}\, .
\end{equation}

\item[D=10, Type~IIA] Taking into account that $\hat{\hat{C}}$ changes
sign under the (now) internal $\Z^{(y)}_{2}$, the eleven-dimensional
transformations become the group

\begin{equation}
SO^{\uparrow}(1,1)_{\rm brane} \times\
SO^{\uparrow}(1,1)_{y} \times\ \Z^{(y)}_{2}\, ,
\end{equation}

\noindent of global symmetries of the equations of motion. The
$SO^{\uparrow}(1,1)$'s act as scalings and the weights and sign
changes under $\Z^{(y)}_{2}$ are summarized in
Table~\ref{tab:d10IIAzero}.

\begin{table}
\begin{center}
\begin{tabular}{||c||c|c|c|c|c|c||}
\hline\hline
& & & & & & \\
Name & $\hat{C}$ & $\hat{g}$ & $\hat{B}^{(1)}$ & $\hat{A^{(1)}}$ &
$e^{\hat{\phi}}$ & $S^{(10)}_{IIA}$ \\
\hline\hline
& & & & & & \\
$SO^{\uparrow}(1,1)_{\rm brane}$ &1&1&1&0&$\frac{1}{2}$&$\frac{3}{2}$ \\
\hline
& & & & & & \\
$SO^{\uparrow}(1,1)_{y}$ &0&1&1&-1&$\frac{3}{2}$&1 \\
\hline
& & & & & & \\
$\Z_{2}^{(y)}$ &$-$&$+$&$+$&$-$&$+$&$+$ \\
& & & & & & \\
\hline\hline
\end{tabular}
\end{center}
\caption{\textit{Weights of the $D=10$ Type~IIA supergravity fields and action
under $SO^{\uparrow}(1,1)_{\rm brane}\times\
SO^{\uparrow}(1,1)_{y}\times\ \Z_{2}^{(y)}$.}}
\label{tab:d10IIAzero}
\end{table}

\item[D=10, Type~IIB] This theory has a manifest $SL(2,\R)$ duality
which in the string frame acts of the fields as follows

\begin{eqnarray}
\hat{\jmath}^{\prime}_{\hat{\mu}\hat{\nu}} & = &
|c\hat{\lambda}+d|\ \hat{\jmath}_{\hat{\mu}\hat{\nu}}\, , \hspace{1.5cm}
\hat{\lambda}^{\prime} = \frac{a\hat{\lambda}+b}{c\hat{\lambda}+d}\, ,
\nonumber \\
& &
\nonumber \\
\left(
\begin{array}{c}
\tilde{\cal{B}}^{(2)\prime}_{\hat{\mu}\hat{\nu}} \\
\tilde{\cal{B}}^{(1)\prime}_{\hat{\mu}\hat{\nu}} \\
\end{array}
\right)
& = &
\left(
\begin{array}{cc}
a & b \\
c & d \\
\end{array}
\right)
\left(
\begin{array}{c}
\tilde{\cal{B}}^{(2)}_{\hat{\mu}\hat{\nu}} \\
\tilde{\cal{B}}^{(1)}_{\hat{\mu}\hat{\nu}} \\
\end{array}
\right)\, ,
\end{eqnarray}

\noindent where $ad-bc=1$ and $\hat{\lambda}=\hat{\ell}+ie^{-\varphi}$.
There are several specially interesting subgroups of $SL(2,\R)$.
One is a $\Z_{2}$ generated by

\begin{equation}
\hat{\lambda}^{\prime}=-1/\lambda\, ,
\hspace{1.2cm}
\hat{\jmath}^{\prime}_{\hat{\mu}\hat{\nu}}=|\lambda|\
\hat{\jmath}_{\hat{\mu}\hat{\nu}}\, ,
\hspace{1.2cm}
\hat{\cal{B}}^{(2)\prime}_{\hat{\mu}\hat{\nu}} =
\hat{\cal{B}}^{(1)}_{\hat{\mu}\hat{\nu}}\, ,
\hspace{.5cm}
\hat{\cal{B}}^{(2)\prime}_{\hat{\mu}\hat{\nu}}=
-\hat{\cal{B}}^{(1)}_{\hat{\mu}\hat{\nu}}\, .
\end{equation}

This transformation inverts the string coupling constant (for
$\hat{\ell}=0$) and that is why it makes sense to identify $SL(2,\R)$
with the $S$-duality group. On the other hand, $\hat{\cal{B}}^{(1)}$ is
a Type~I field, whose origin is the elementary excitations of the
string, but $\hat{\cal{B}}^{(2)}$ is a Ramond-Ramond-type
field, whose origin is in solitonic modes on the worldsheet. Therefore,
this transformation has also an ``electric-magnetic'' side from the
worldsheet point of view.

Another subgroup is a scaling $\widetilde{SO}^{\uparrow}(1,1)_{y}$ given
in Table~\ref{tab:d10IIBzero}.  It can be obtained from the
$SO^{\uparrow}(1,1)_{y}$ of Type~IIA using Type~II Buscher duality
\cite{Be2}. Using it, we have also translated
$SO^{\uparrow}(1,1)_{\rm brane}\times\ \Z_{y}$ to the Type~IIB language.
the results are given also in Table~\ref{tab:d10IIBzero}.

\begin{table}
\begin{center}
\begin{tabular}{||c||c|c|c|c|c|c|c||}
\hline\hline
& & & & & & & \\
Name & $\hat{D}$ & $\hat{\jmath}$ & $\hat{\cal B}^{(1)}$ &
$\hat{\cal B}^{(2)}$ & $\hat{\ell}$ & $e^{\hat{\varphi}}$ &
$S^{(10)}_{IIB}$ \\
\hline\hline
& & & & & & & \\
$\widetilde{SO}^{\uparrow}(1,1)_{\rm brane}$ &1&1&1&0&-1&1&2 \\
\hline
& & & & & & & \\
$\widetilde{SO}^{\uparrow}(1,1)_{y}$ &0&1&1&-1&-2&2&0 \\
\hline
& & & & & & & \\
$\tilde{\Z}_{2}^{(y)}$ &$-$&$+$&$+$&$-$&$-$&$+$&$+$ \\
\hline\hline
\end{tabular}
\end{center}
\caption{\textit{Weights of the $D=10$ Type~IIB supergravity fields and
($\hat{F}(\hat{D})=0$ truncated) action under
$\widetilde{SO}^{\uparrow}(1,1)_{\rm brane}\times\
\widetilde{SO}^{\uparrow}(1,1)_{y}\times\ \tilde{\Z}_{2}^{(y)}$.}}
\label{tab:d10IIBzero}
\end{table}

The total group on global symmetries of the equations of motion is, then

\begin{equation}
GL(2,\R)=SL(2,\R)\times\ \widetilde{SO}^{\uparrow}(1,1)_{\rm brane}
\times\ \tilde{\Z}_{2}^{(y)}\, .
\end{equation}

We would like to remark that this is exactly the global symmetry group
that one would expect in a ten-dimensional theory that has been obtained
by dimensional reduction from a twelve-dimensional theory with no
global symmetries whatsoever.

\item[D=10, Type~I] Truncating any of the Type~II theories by setting
the Ramond-Ramond fields to zero we obtain the symmetries of the
Type~I theory. These are

\begin{equation}
SO^{\uparrow}(1,1)_{\rm brane} \times\
SO^{\uparrow}(1,1)_{y} \times\ \Z^{(A)}_{2}\, .
\end{equation}

The $SO^{\uparrow}(1,1)_{y}$ group is the same as in Type~IIA and the
same as $\widetilde{SO}^{\uparrow}(1,1)_{y}$ in Type~IIB, and is the
only subgroup that remains of $SL(2,\R)$.

\end{description}

\subsection{Two isometries}

\begin{description}

\item[D=11] Upon dimensional reduction to nine dimensions, the global
symmetry group that we expect is

\begin{equation}
GL(2,\R) \times\ SO^{\uparrow}(1,1)_{\rm brane} =SL(2,\R)\times\
SO^{\uparrow}(1,1)\times\ SO^{\uparrow}(1,1)_{\rm brane} \times\Z_{2}\, .
\end{equation}

\item[D=10, Type~IIA and B] In presence of an isometry (in ten
dimensions), the Type~IIA and Type~IIB theories are related by
Type~II duality \cite{Be3}. There are other global symmetries which are
not covariant from the ten-dimensional point of view. They are
become covariant when we rewrite the theories in nine-dimensional
language and so we will discuss them below.

\item[D=10, Type~I] Upon truncation of the Ramond-Ramond-type fields
both Type~II theories become the Type~I theory and the Type~II
duality that related them becomes the $\Z^{(B)}_{2}$ Buscher duality
that we also discuss below in nine dimensions.

\item[D=9, Type~II] In nine dimensions there is a single Type~II
theory whose global symmetry group is the one we expected:

\begin{equation}
SL(2,\R) \times\ SO^{\uparrow}(1,1)_{x+y} \times\
SO^{\uparrow}(1,1)_{\rm brane} \times\ \Z^{(x)}_{2}\, .
\end{equation}

This $SL(2,\R)$ group is a symmetry of the action.  From the Type~IIB
point of view it is the manifest $SL(2,\R)$ symmetry of the original
theory and from the point of view of the Type~IIA is part of the
symmetry predicted in eleven dimensions.  It contains one particular
subgroup of scalings: $SO^{\uparrow}(1,1)_{x-y}$ corresponding to the
eleven-dimensional g.c.t.~$x\rightarrow e^{\alpha}x\, ,y \rightarrow
e^{-\alpha}y$. $SO^{\uparrow}(1,1)_{x+y}$ scales the fields and the
action and corresponds to the eleven-dimensional g.c.t.~$x\rightarrow
e^{\alpha}x\, ,y \rightarrow e^{\alpha}y$.  Combining it with
$SO^{\uparrow}(1,1)_{\rm brane}$, a second scaling symmetry of the action
can be obtained.  Finally, $\Z_{2}^{(x)}$ corresponds to improper g.c.t.s in
the internal space, for instance $x\rightarrow -x$ (up to $SL(2,\R)$
rotations).  The weights of the different nine-dimensional fields are
summarized in Table~\ref{tab:d9II}.

\begin{table}
\begin{center}
\begin{tabular}{||c||c|c|c|c|c|c|c|c|c|c|c||}
\hline\hline
& & & & & & & & & & & \\
Name & $C$ & $g$ & $B^{(1)}$ & $B^{(2)}$ & $A^{(1)}$ & $A^{(2)}$ & $B$ &
$k$ & $\ell$ & $e^{\phi}$ & $S^{(9)}_{\rm II-red}$ \\
\hline\hline
& & & & & & & & & & & \\
$SO^{\uparrow}(1,1)_{\rm brane}$
&1&1&1&1&0&0&1&$\frac{1}{2}$&0&$\frac{1}{4}$&3 \\
\hline
& & & & & & & & & & & \\
$SO^{\uparrow}(1,1)_{x-y}$
&0&1&1&-1&-1&1&0&$-\frac{1}{2}$&-2&$\frac{7}{4}$&0 \\
\hline
& & & & & & & & & & & \\
$SO^{\uparrow}(1,1)_{x+y}$
&0&1&1&1&-1&-1&2&$\frac{3}{2}$&0&$\frac{3}{4}$&2 \\
\hline
& & & & & & & & & & & \\
$\Z_{2}^{(x)}$
&$-$&$+$&$-$&$+$&$+$&$-$&$+$&$+$&$-$&$+$&$+$ \\
\hline\hline
\end{tabular}
\end{center}
\caption{\textit{Weights of the $D=9$ Type~II supergravity fields and
action under $SL(2,\R) \times\ SO^{\uparrow}(1,1)_{x+y} \times\
SO^{\uparrow}(1,1)_{\rm brane} \times\ \Z^{(x)}_{2}$.}}
\label{tab:d9II}
\end{table}

\item[D=9, Type~I] After truncation of the Type~II theory, two
interesting and opposite phenomena take place: the breaking of the
Type~I $SL(2,\R)$ to just $SO^{\uparrow}(1,1)_{x-y}$, and a
discrete symmetry enhancement from $\Z_{2}^{(x)}$ to $D_{4}$, due to the
appearance of two new $\Z_{2}$'s: $\Z_{2}^{(A)}$ and $\Z_{2}^{(B)}$.
The appearance of $\Z_{2}^{(A)}$ is related to the disappearance of the
topological term in the action. The appearance of $\Z_{2}^{(B)}$ is
more subtle and was discussed in Ref.~\cite{Be3}. In
Table~\ref{tab:weights-I} the weights of the nine-dimensional Type~I
fields under certain combination of these symmetries are given.

\end{description}


\end{document}